\documentclass[letterpaper,prc,superscriptaddress,showpacs,twocolumn,amssymb,amsmath,amsfonts,aps]{revtex4-1}
\usepackage[pdftex]{graphicx}
\usepackage{dcolumn}  
\usepackage{longtable}
\usepackage{bm}       
\usepackage{subfigure}
\usepackage{rotating}
\usepackage{epstopdf}
\newcommand\captionof[1]{\def\@captype{#1}\caption}
\begin{document}
\author{C.~A. Meyer}
\author{Y.~Van~Haarlem}
\affiliation{Carnegie Mellon University, Pittsburgh, PA 15213}

\date{\today}
\title{The Status of Exotic-quantum-number Mesons}
\begin{abstract}
The search for mesons with non-quark-antiquark (exotic) quantum numbers has gone
on for nearly thirty years. There currently is experimental evidence of three isospin one
states, the $\pi_{1}(1400)$, the $\pi_{1}(1600)$ and the $\pi_{1}(2015)$. For all of these
states, there are questions about their identification, and even if some of
them exist. In this article, we will review both the theoretical work and the experimental
evidence associated with these exotic quantum number states. We find that the $\pi_{1}(1600)$
could be the lightest exotic quantum number hybrid meson, but observations of other members
of the nonet would be useful. 
\end{abstract}
\pacs{14.40.-n,14-40.Rt,13.25.-k}
\maketitle
\section{\label{sec:intro}INTRODUCTION}
The quark model describes mesons as bound states of quarks and antiquarks ($q\bar{q}$),
much akin to positronium ($e^{+}e^{-}$).  As described in Section~\ref{sec:quark-model},
mesons have well-defined quantum numbers: total spin $J$, parity $P$, and C-parity $C$,
represented as $J^{PC}$. The allowed $J^{PC}$ quantum numbers for orbital angular momentum, $L$, 
smaller than three are given in Table~\ref{tab:qns}. Interestingly, for $J$ smaller than $3$, 
all allowed $J^{PC}$ except $2^{--}$~\cite{amsler08} have been observed by experiments.
From the allowed quantum numbers in Table~\ref{tab:qns}, there are several missing 
combinations: $0^{--}$, $0^{+-}$, $1^{-+}$ and $2^{+-}$. These are not possible 
for simple $q\bar{q}$ systems and are known as ``exotic'' quantum numbers.
Observation of states with exotic quantum numbers has been of great experimental interest 
as it would be clear evidence for mesons beyond the simple $q\bar{q}$ picture. 
\begin{table}[h!]\centering
\begin{tabular}{ccr|ccr|ccr} \hline\hline
$L$ & $S$ & $J^{PC}$  & $L$ & $S$ & $J^{PC}$  & $L$ & $S$ & $J^{PC}$ \\ \hline
$0$ & $0$ & $0^{-+}$ &
$1$ & $0$ & $1^{+-}$ &
$2$ & $0$ & $2^{-+}$ \\
$0$ & $1$ & $1^{--}$ &
$1$ & $1$ & $0^{++}$ &
$2$ & $1$ & $1^{--}$ \\
       &        &               &
$1$ & $1$ & $1^{++}$ & 
$2$ & $1$ & $2^{--}$ \\
       &        &               &
$1$ & $1$ & $2^{++}$ &
$2$ & $1$ & $3^{--}$ \\
\hline\hline
\end{tabular}
\caption[]{\label{tab:qns}The allowed $J^{PC}$ quantum numbers for $q\bar{q}$ systems.}
\end{table}

Moving beyond the simple quark-model picture of mesons, there have been predictions 
for states with these exotic quantum numbers. The most well known are $q\bar{q}$ states
in which the gluons binding the system can contribute directly to the quantum numbers of
the meson. However, other candidates include multi-quark states ($q\bar{q}q\bar{q}$) and
states containing only gluons (glueballs). Early bag-model calculations~\cite{Barnes:1982zs} 
referred to states with $q\bar{q}$ and gluons as ``hermaphrodite mesons'', and predicted that 
the lightest nonet ($J^{PC}=1^{-+}$) might have masses near $1$~GeV as well as distinctive decay 
modes. They might also be relatively stable, and thus observable. While the name hermaphorodite 
did not survive, what are now known as ``hybrid mesons'' have become a very interesting theoretical 
and experimental topic and the status of these states, with particular emphasis on the exotic-quantum
number ones is the topic of this article. More information on meson spectroscopy in general can be 
found in a recent review by Klempt and Zaitsev~\cite{klempt-07}. Similarly, a recent review on 
the related topic of glueballs can be found in reference~\cite{crede-09}.

\section{Theoretical Expectations for Hybrid Mesons}
\subsection{\label{sec:quark-model}Mesons in The Quark Model}
In the quark model, mesons are bound states of quarks and antiquarks ($q\bar{q}$). 
The quantum numbers of such fermion-antifermion systems are functions of the 
total spin, $S$, of the quark-antiquark system, and the relative orbital angular 
momentum, $L$, between them. The spin $S$ and angular momentum $L$ combine to 
yield the total spin
\begin{eqnarray}
J & = & L \oplus S \, ,
\end{eqnarray}
where $L$ and $S$ add as two angular momentums. 

Parity is the result of a mirror reflection of the wave function, taking $\vec{r}$ into $-\vec{r}$.
It can be written as
\begin{eqnarray}
P\left[\psi(\vec{r})\right] & = \psi(-\vec{r}) & = \eta_{P}\psi(\vec{r}) \, ,
\end{eqnarray}
where $\eta_{P}$ is the eigenvalue of parity. As application of parity twice must return the 
original state, $\eta_{P}=\pm 1$. In spherical coordinates, the parity operation reduces to the 
reflection of a $Y_{lm}$ function,
\begin{eqnarray}
Y_{lm}(\pi-\theta,\pi+\phi) & = & (-1)^{l}Y_{lm}(\theta,\phi) \, .
\end{eqnarray} 
From this, we conclude that $\eta_{P}=(-1)^{l}$. 

For a $q\bar{q}$ system, the intrinsic parity of the antiquark is opposite to that of the 
quark, which yields the total parity of a $q\bar{q}$ system as 
\begin{eqnarray}
P(q\bar{q}) & = & -(-1)^{L} \, .
\end{eqnarray}

Charge conjugation, $C$, is the result of a transformation that takes a particle into its antiparticle. For
a $q\bar{q}$ system, only electrically-neutral states can be eigenstates of $C$. In order to
determine the eigenvalues of $C$ ($\eta_{C}$), we need to consider a wave function that includes
both spatial and spin information
\begin{eqnarray}
\Psi(\vec{r},\vec{s}) & = & R(r) Y_{lm}(\theta,\phi) \chi(\vec{s}) \, .
\end{eqnarray}

As an example, we consider a $u\bar{u}$ system, the $C$ operator acting on this reverses the 
meaning of $u$ and $\bar{u}$. This has the effect of mapping the vector $\vec{r}$ to 
the $u$ quark into $-\vec{r}$. Thus, following the arguments for parity, the spatial part 
of $C$ yields a factor of $(-1)^{L}$. The spin wave function also reverse the two individual
spins. For a symmetric $\chi$, we get a factor of $1$, while for an antisymmetric $\chi$,
we get a factor of $-1$. For two spin $\frac{1}{2}$ particles, the $S=0$ singlet is antisymmetric,
while the $S=1$ triplet is symmetric. Combining all of this, we find that the C-parity of 
(a neutral) $q\bar{q}$ system is 
\begin{eqnarray}
C(q\bar{q}) & = & (-1)^{L+S} \, .
\end{eqnarray} 

Because $C$-parity is only defined for neutral states, it is useful to extend this to the 
more general $G$-parity which can be used to describe all $q\bar{q}$ states, independent of
charge. For isovector states ($I=1$), $C$ would transform a charged member into the oppositely
charged state (\emph{e.g.} $\pi^{+}\to\pi^{-}$). In order to transform this back to the original 
charge, we would need to perform a rotation in isospin ($\pi^{-}\to\pi^{+}$). For a state of whose 
neutral member has $C$-parity $C$, and whose total isospin is $I$, the $G$-parity is defined to be 
\begin{eqnarray}
G & = & C\cdot (-1)^{I} \, , 
\end{eqnarray}
which can be generalized to 
\begin{eqnarray}
\label{eq:g-parity}
G(q\bar{q}) & = & (-1)^{L+S+I} \, .
\end{eqnarray}
The latter is valid for all of the $I=0$ and $I=1$ members of a nonet. 
This leads to mesons having well defined quantum numbers: total angular momentum, 
$J$, isospin, $I$, parity $P$, C-parity, $C$, and G-parity, $G$. These are represented as 
$(I^{G}) J^{PC}$, or simply $J^{PC}$ for short. For the case of $L=0$ and $S=0$, we have 
$J^{PC}=0^{-+}$, while for $L=0$ and $S=1$, $J^{PC}=1^{--}$. The allowed quantum numbers 
for  $L$ smaller than three are given in Table~\ref{tab:qns}. 

\subsection{Notation and Quantum Numbers of Hybrids}
The notation for hybrid mesons we use is that from the Particle Data Group (PDG)~\cite{amsler08}. 
In the PDG notation, the parity and charge conjugation determine the name of the hybrid, which
is taken as the name of the normal meson of the same $J^{PC}$ and isospin. The total spin is then 
used as a subscript to the name. While various models predict different nonets of hybrid mesons, 
the largest number of nonets is from the flux-tube model (see Section~\ref{sec:models}). For completeness, we list all of these as well as 
their PDG names in Table~\ref{tab:hybrid-names}.  The first entry is the isospin one ($I=1$) state. 
The second and third are those with isospin equal to zero ($I=0$)  and the fourth is the kaon-like
state with isospin one-half ($I=\frac{1}{2}$). 
In the case of the $I=0$ states, the first is taken as the mostly $u\bar{u}$ and $d\bar{d}$ state 
(so-called $n\bar{n}$), while the second is mostly $s\bar{s}$. For the $I=0$ states, $C$-parity 
is well defined, but for $I=1$, only the neutral member can have a defined $C$-parity. However, 
the more general $G$-parity can be used to describe all of the $I=1$ members (see 
equation~\ref{eq:g-parity}).  Thus, the $G$-parity can be used to identify exotic quantum numbers, 
even for charged $I=1$ members of a nonet. For the case of the kaon-like states, neither $C$-parity 
nor $G$-parity is defined. Thus, the $I=\frac{1}{2}$ members of a nonet can not have 
explicitly-exotic quantum numbers.
\begin{table}[h]\centering
\begin{tabular}{cccccccc} \hline\hline
\multicolumn{2}{c}{QNs} & \multicolumn{6}{c}{Names} \\ 
 $J^{PC}$    &  $(I^{G})$   &   &  $(I^{G})$ & & & $(I)$ & \\ \hline
$1^{++}$ & $(1^{-})$& $a_{1}$        & $(0^{+})$ & $f_{1}$       & $f^{\prime}_{1}$ &$(\frac{1}{2})$  &$K_{1}$   \\
$1^{--}$ & $(1^{+})$ & $\rho_{1}$   & $(0^{-})$ &$\omega_{1}$ & $\phi_{1}$  &$(\frac{1}{2})$  &$K^{*}_{1}$ \\ \hline
$0^{-+}$ & $(1^{-})$ &  $\pi_{0}$    & $(0^{+})$ &$\eta_{0}$ & $\eta^{\prime}_{0}$ &$(\frac{1}{2})$   &$K_{0}$   \\
$\mathbf{1^{-+}}$ & $\mathbf{(1^{-})}$ & $\mathbf{\pi_{1}}$     & $\mathbf{(0^{+})}$ &$\mathbf{\eta_{1}}$ & $\mathbf{\eta^{\prime}_{1}}$ &$(\frac{1}{2})$   &$K^{*}_{1}$ \\
$2^{-+}$ & $(1^{-})$ & $\pi_{2}$     & $(0^{+})$ &$\eta_{2}$ & $\eta^{\prime}_{2}$ &$(\frac{1}{2})$   &$K_{2}$  \\ \hline
$\mathbf{0^{+-}}$ & $\mathbf{(1^{+})}$ &  $\mathbf{b_{0}}$       & $\mathbf{(0^{-})}$&$\mathbf{h_{0}}$     & $\mathbf{h^{\prime}_{0}}$    &$(\frac{1}{2})$     &$K^{*}_{0}$ \\
$1^{+-}$ & $(1^{+})$ & $b_{1}$        & $(0^{-})$&$h_{1}$     & $h^{\prime}_{1}$     &$(\frac{1}{2})$    &$K_{1}$  \\
$\mathbf{2^{+-}}$ & $\mathbf{(1^{+})}$ & $\mathbf{b_{2}}$        & $\mathbf{(0^{-})}$&$\mathbf{h_{2}}$     & $\mathbf{h^{\prime}_{2}}$     &$(\frac{1}{2})$    &$K^{*}_{2}$ \\ \hline \hline
\end{tabular}
\caption[]{\label{tab:hybrid-names}The naming scheme for hybrid mesons. The first
state listed for a given quantum number is the isospin one state. The second state is
the isospin zero state that is mostly $u$ and $d$ quarks ($n\bar{n}$), while the third
name is for the mostly $s\bar{s}$ isospin zero state. Note that for the kaons, the $C$-
and $G$-parity are not defined. Kaons cannot not have manifestly exotic quantum 
numbers. States that have exotic quantum numbers are shown in bold.}
\end{table}

In Table~\ref{tab:g-parity} we show the $J^{P}$ of the three exotic $I=1$ mesons from 
Table~\ref{tab:hybrid-names}. We also show the normal ($q\bar{q}$) meson of the same $J^{P}$ and
the $I^{G}$ quantum numbers for these states. The exotic mesons have the opposite $G$-parity 
relative to the normal meson. This provides a simple mechanism for identifying if a charged 
$I=1$ state has exotic quantum numbers.
\begin{table}[h!]\centering
\begin{tabular}{ccccc} \hline\hline
$J^{P}$   & \multicolumn{2}{c}{normal meson} &  \multicolumn{2}{c}{exotic meson} \\
             & name & $(I^{G})$ & name & $(I^{G})$ \\ \hline
$0^{+}$  & $a_{0}$  & $(1^{-})$ &  $b_{0}$ & $(1^{+})$ \\
$1^{-}$  &  $\rho$ & $(1^{+})$ &  $\pi_{1}$ & $(1^{-})$ \\
$2^{+}$  &  $a_{2}$ & $(1^{-})$ &  $b_{2}$ & $(1^{+})$ \\ 
\hline\hline
\end{tabular}
\caption[]{\label{tab:g-parity}The $J^{P}$ and $I^{G}$ quantum numbers for the exotic mesons
and the normal mesons of the same $J^{P}$.}
\end{table}
\subsection{\label{sec:models}Model Predictions}
The first predictions for exotic quantum number mesons came from calculations 
in the Bag model~\cite{Jaffe-76,Vainshtein:1978nn}. In this model, boundary conditions
are placed on quarks and gluons confined inside a bag. A hybrid meson is formed by
combining a $q\bar{q}$ system (with spin $0$ or $1$) with a transverse-electric (TE) gluon
($J^{PC}=1^{+-}$). This yields four nonets of hybrid mesons with quantum numbers
$J^{PC}=1^{--}$, $0^{-+}$, $1^{-+}$ and $2^{-+}$. These four nonets are roughly degenerate
in mass and early calculations predicted the mass of a $1^{-+}$ to be in the range of $1.2$ 
to $1.4$~GeV~\cite{Barnes:1982tx,Chanowitz:1982qj}. In the bag model, the 
transverse-magnetic gluon is of higher mass. It has $J^{PC}=1^{-+}$ and combined with
the same $S=0$ and $S=1$ $q\bar{q}$ systems yield four additional nonets with
$J^{PC}=1^{++}$, $0^{+-}$, $1^{+-}$ and $2^{+-}$. These would presumably be heavier
than the nonets built with the TE gluon.

Another method that has been used to predict the hybrid masses are ``QCD spectral 
sum rules''  (QSSR). Using QSSR, one examines a two-point correlator of appropriate 
field operators from QCD and produces a sum rule by equating a dispersion relation for 
the correlator to an operator product expansion. QSSR calculations initially found a $1^{-+}$ 
state near $1$~GeV~\cite{Balitsky:1982ps,Latorre-87}. A $0^{--}$ state was also predicted
around $3.8$~GeV in mass~\cite{Latorre-87}. Newer calculations~\cite{Narison-00} tend to 
favor a $1^{-+}$ hybrid mass in the range of $1.6$ to $2.1$~GeV, and favor the $\pi_{1}(1600)$ 
(see Section~\ref{sec:pi1_1600}) as the  lightest exotic hybrid. Recently,  
Narison~\cite{Narison-09} looked at the calculations for  $J^{PC}=1^{-+}$ states with particular
emphasis in understanding differences in the results between QSSR and Lattice QCD calculations
(see Section~\ref{sec:lqcd}).  He found that the $\pi_{1}(1400)$ and $\pi_{1}(1600)$ may be 
consistent with 4-quark states, while QSSR are consistent with the $\pi_{1}(2015)$ (see 
Section~\ref{sec:pi1_2015}) being the lightest hybrid meson.

The formation of flux tubes was first introduced in the 1970's
by Yoichiro Nambu~\cite{nambu-70,nambu-76} to explain the observed linear 
Regge trajectories---the linear dependence of mass squared, $m^{2}$, of 
hadrons on their spin, $J$.  This linear dependence results if one assumes 
that mass-less quarks are tied to the ends of a relativistic string with 
constant mass (energy) per length and the system rotating about its center.
The linear $m^{2}$ versus $J$ dependence only arises when the mass density 
per length is constant, which is equivalent to a linear potential.

In the heavy-quark sector, lattice QCD~\cite{bali-98} calculations show a distribution
of the gluonic field (action density) which is mostly confined to the region between the
quark and the antiquark. A picture which is very similar to that inspired by the ``flux-tube 
model''.  Within the flux-tube model~\cite{isgur-85a,isgur-85b}, one can view 
hybrids as mesons with angular momentum in the flux tube. Naively, 
one can imagine two degenerate excitations, one with the tube going 
clockwise and one counter clockwise. It is possible to write linear 
combinations of these that have definite spin, parity and C-parity.
For the case of one unit of angular momentum in the tube, the flux tube behaves as 
if it has quantum numbers $J^{PC}=1^{+-}$ or $1^{-+}$. The basic quantum numbers 
of hybrids are obtained by adding the tube's quantum numbers to that of 
the underlying  meson. 

In the flux-tube model, the tube carries angular momentum, $m$, which then leads to 
specific predictions for the product of $C$-parity and parity ($CP$). For $m=0$,  
one has $CP=(-1)^{S+1}$, while for the first excited states, ($m=1$), we find that 
$CP=(-1)^{S}$. The excitations are then built on top of the $s$-wave mesons, ($L=0$), 
where the total spin can be either $S=0$ or $S=1$. For the case of $m=0$, we find 
$CP$ as follows, 
\begin{eqnarray*}
(m=0) & \left . 
\begin{array}{cc} S=0 & 0^{-+} \\ S=1 & 1^{--} \end{array}
\right \} & \begin{array}{c} (-1)^{L+1}(-1)^{S+L}=(-1)^{S+1} \\
\mathrm{Normal\, Mesons} \end{array} \\
\end{eqnarray*}
which are the quantum numbers of the normal, $q\bar{q}$, mesons as discussed in 
Section~\ref{sec:quark-model}. For the case of $m=1$, where we have one unit of angular
momentum in the flux tube, we find the following $J^{PC}$ quantum numbers
\begin{eqnarray*}
(m=1) & \left . \begin{array}{cc} S=0 & 0^{-+} \\ S=1 & 1^{--} \end{array}
\right \} & \begin{array}{c} 1^{++}, 1^{--} \\
0^{-+}, \mathbf{0^{+-}}, \mathbf{1^{-+}}, 1^{+-}, 2^{-+}, \mathbf{2^{+-}} \, .
\end{array}
\end{eqnarray*}
The resulting quantum numbers are obtained by adding both $1^{+-}$ and 
$1^{-+}$ to the underlying $q\bar{q}$ quantum numbers ($0^{-+}$ and $1^{--}$).

From the two $L=0$ meson nonets, we expect eight hybrid nonets, (72 new mesons!).
Two of these nonets arise from the $q\bar{q}$ in an $S=0$ (singlet) state, while six
arise for the $q\bar{q}$ in the $S=1$ (triplet) state. Of the six states built on the triplet
$q\bar{q}$, three have exotic quantum numbers (as indicated in bold above).

In the picture presented by the flux-tube model, the hybrids are no different than 
other excitations of the $q\bar{q}$ states. In addition to ``orbital'' and ``radial''
excitations, we also need to consider ``gluonic'' excitations. Thus, the flux-tube model
predicts eight nonets of hybrid mesons ($0^{+-}$, $0^{-+}$, $1^{++}$, $1^{--}$, $1^{-+}$, $1^{+-}$,
$2^{-+}$ and $2^{+-}$). The model also predicts that all eight nonets are degenerate in
mass, with masses expected near $1.9$~GeV~\cite{isgur-85b}.

An alternate approach to calculating properties of hybrid mesons comes
from the effective QCD Coulomb-gauge Hamiltonian. Here, Foch states for hadrons are
constructed from the vacuum as well as quark and gluon operators. In this model, the
lightest hybrid nonets are $J^{PC}=1^{+-}$, $0^{++}$, $1^{++}$ and $2^{++}$, none of which
are exotic. The first excitation of these ($L=1$), yields the nonets $1^{-+}$, $3^{-+}$
and $0^{--}$, all of which are exotic~\cite{Cotanch-01,Cotanch:2006wv}. In this model, the $1^{-+}$ is the
lightest exotic quantum number hybrid, with a mass in the range of $2.1$ to $2.3$~GeV.
Predictions are also made for the lightest $c\bar{c}$ exotic hybrid, which is found in the
range of $4.1$ to $4.3$~GeV.

In Table~\ref{tab:model-masses} are presented a summary of the mass predictions for the
various model calculations for hybrid meson masses.
\begin{table}[h!]\centering
\begin{tabular}{clc} \hline\hline
 Mass (GeV)      & Model & Reference \\ \hline
 $1.0$- $1.4$  & Bag Model & \cite{Jaffe-76,Vainshtein:1978nn,Barnes:1982zs} \\
 $1.0$-$1.9$  & QSSR &\cite{Balitsky:1982ps,Latorre-87,Narison-00,Narison-09} \\
 $1.8$-$2.0$  & Flux Tube &\cite{isgur-85b} \\
 $2.1$-$2.3$  & Hamiltonian &\cite{Cotanch-01} \\
\hline\hline
\end{tabular}
\caption[]{\label{tab:model-masses}Mass predictions for hybrid mesons from various models.}
\end{table}

\subsection{\label{sec:lqcd}Lattice Predictions}
Lattice QCD (LQCD) calculations may provide the most accurate estimate to the masses of
hybrid mesons. While these calculations have progressively gotten better,
they are still limited by a number of systematic effects. Currently, the most significant 
of these is related to the mass of the light quarks used in the calculations.
This is typically parametrized as the pion mass, and extrapolations need
to be made to reach the physical pion mass. This is often made as a linear
approximation, which may not be accurate. In addition, as the the quark mass 
becomes lighter, two-meson decay channels become possible. These may 
distort the resulting spectrum.  

Most calculations have been performed with what is effectively the strange-quark mass. 
However, it may not be safe to assume that this is the mass of the $s\bar{s}$ member of
the nonet, and one needs to be aware of the approximations made to move the estimate 
to the $u\bar{u}/d\bar{d}$ mass.  The bottom line is that no one would be surprised if the 
true hybrid masses differed by several hundred MeV from the best predictions.
\begin{table}[h!]\centering
\begin{tabular}{ccc}\hline\hline
Author &
\multicolumn{2}{c}{$1^{-+}$ Mass  (GeV/c$^{2}$)} \\
Collab.  & $u\bar{u}/d\bar{d}$ & $s\bar{s}$ \\
\hline
UKQCD~\cite{lacock-97}  & $1.87\pm 0.20$ & $2.0\pm 0.2$ \\
MILC~\cite{bernard-97} & $1.97\pm 0.09\pm 0.30$ 
                                       & $2.170\pm 0.080 \pm 0.30$ \\
SESAM~\cite{lacock-98}  & $1.9\pm 0.20$        &  \\
MILC~\cite{bernard-99} & $2.11\pm 0.10 \pm (sys)$       &  \\
Mei~\cite{Mei:2002ip}   & $2.013\pm 0.026\pm 0.071$ & \\
Hedditch~\cite{hedditch-05} & $1.74\pm 0.25$ & \\
Bernard~\cite{bernard-04} & $1.792\pm 0.139$ 
                                                   & $2.100\pm 0.120$  \\%

McNeile~\cite{McNeile:2006bz} & $2.09\pm 0.1$ & \\
\hline\hline
\end{tabular}
\caption[Lattice Predictions for Hybrids]{\label{tab:lattice-hybrid}
Recent results for the light-quark  $1^{-+}$ hybrid meson masses.}
\end{table}

While the flux-tube model (see Section~\ref{sec:models} predicts that the lightest eight
nonets of hybrid mesons are degenerate in mass at about $1.9$~GeV, LQCD calculations
consistently show that the $J^{PC}=1^{-+}$ nonet is the lightest. Predictions for the 
mass of this state have varied from $1.8$ to $2.1$~GeV, with an average about in the 
middle of these.  Table~\ref{tab:lattice-hybrid} shows a number of these predictions 
made over the last several years. Most of these~\cite{lacock-97,bernard-97,lacock-98,
bernard-99, Mei:2002ip,hedditch-05} were made in the quenched approximation 
(no $q\bar{q}$ loops allowed in the quenched calculation), while newer calculations~\cite{bernard-04,
McNeile:2006bz,Dudek:2009qf,Dudek:2010wm} are dynamic (not quenched). 

\begin{figure}[h!]\centering
\includegraphics[width=0.45\textwidth]{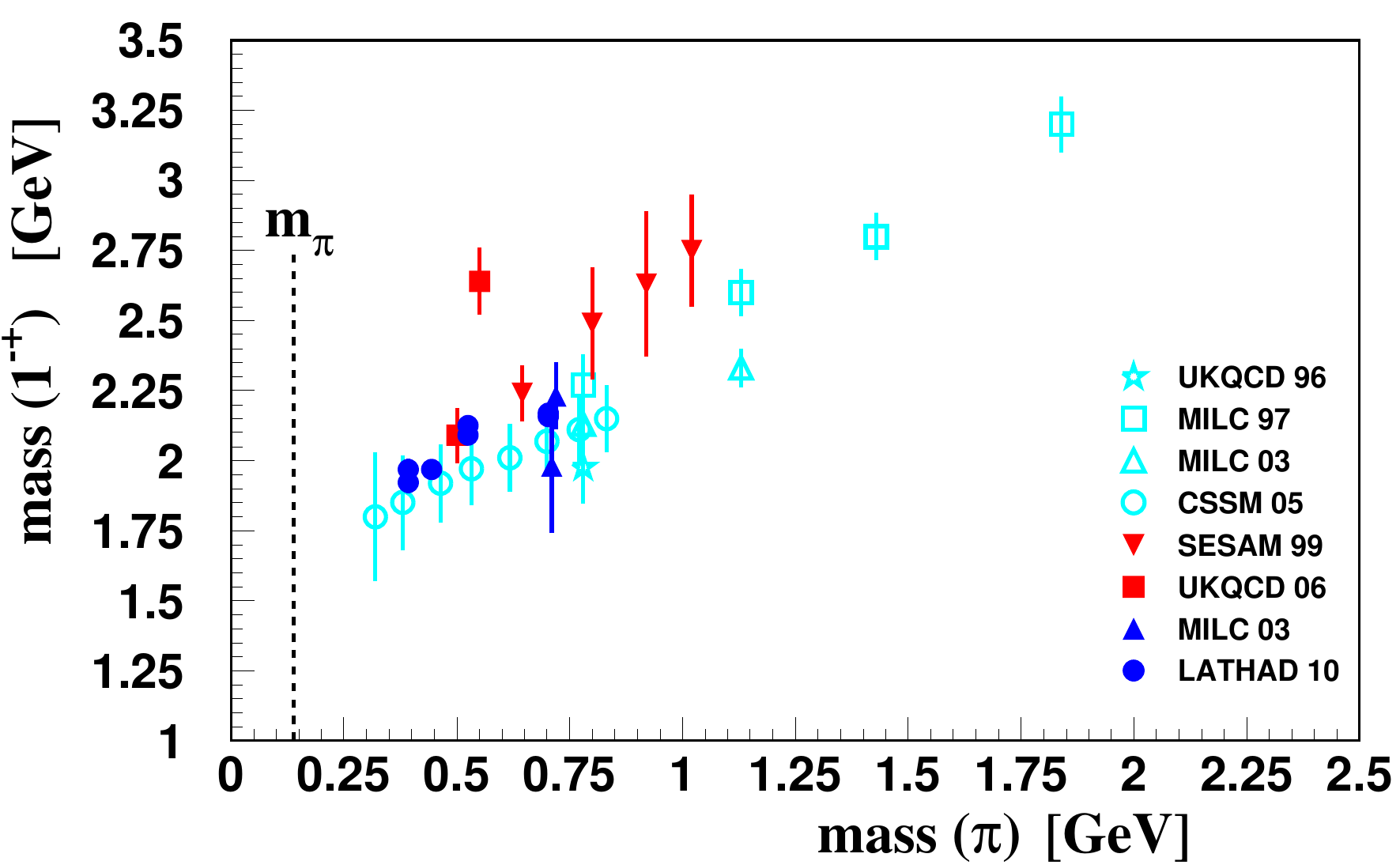}
\caption[]{\label{fig:exotic_masses}(Color on line.) The mass of the $J^{PC}=1^{-+}$ exotic hybrid
as a function of the pion mass from lattice calculations. The open (cyan) symbols correspond to 
quenched calculations, while the solid (red and blue) symbols are dynamic (unquenched)
calculations: open (cyan) star~\cite{lacock-97}, open (cyan) squares~\cite{bernard-97}, open (cyan) upright
triangles~\cite{bernard-04}, open (cyan) circles~\cite{hedditch-05}, solid (red) 
downward triangles~\cite{lacock-98}, solid (red) squares~\cite{McNeile:2006bz}, solid (blue) 
upright triangles~\cite{bernard-04} and solid (blue) circles~\cite{Dudek:2010wm}.}
\end{figure}

However, the masses in Table~\ref{tab:lattice-hybrid} may not be the best approximations to 
the hybrid masses. It has been noted~\cite{dudek-priv} that Table~\ref{tab:lattice-hybrid} is not a very 
useful way of displaying the results. Rather, the mass needs to be correlated with the light-quark
mass used in the calculation. This is usually represented as the pion mass. In 
Figure~\ref{fig:exotic_masses} are shown the predictions from the same groups as a function 
of the pion masses used in their calculations. In order to obtain the hybrid mass, one needs to 
extrapolate to the physical pion mass. 

There are fewer predictions for the masses of the other exotic-quantum 
number states. Bernard~\cite{bernard-97} calculated the splitting between
the $0^{+-}$ and the $1^{-+}$ state to be about $0.2$~GeV with large
errors. A later calculation using a clover action~\cite{bernard-99} 
found a splitting of $0.270 \pm 0.2$~GeV. The SESAM collaboration~\cite{lacock-98} has one such 
calculation, the results of which are shown in Table~\ref{tab:other1}.
\begin{table}[h!]\centering
\begin{tabular}{ccc} \hline\hline
Multiplet & $J^{PC}$ & Mass \\ \hline
$\pi_{1}$ & $1^{-+}$ & $1.9\pm 0.2 \, GeV/c^{2}$ \\
$b_{2}$   & $2^{+-}$ & $2.0\pm 1.1 \, GeV/c^{2}$ \\
$b_{0}$   & $0^{+-}$ & $2.3\pm 0.6 \, GeV/c^{2}$ \\
\hline\hline
\end{tabular}
\caption[]{\label{tab:other1}Estimates of the masses of exotic quantum number
hybrids~\cite{lacock-98}.}
\end{table}
\begin{figure*}[t!]\centering
\includegraphics[width=0.75\textwidth]{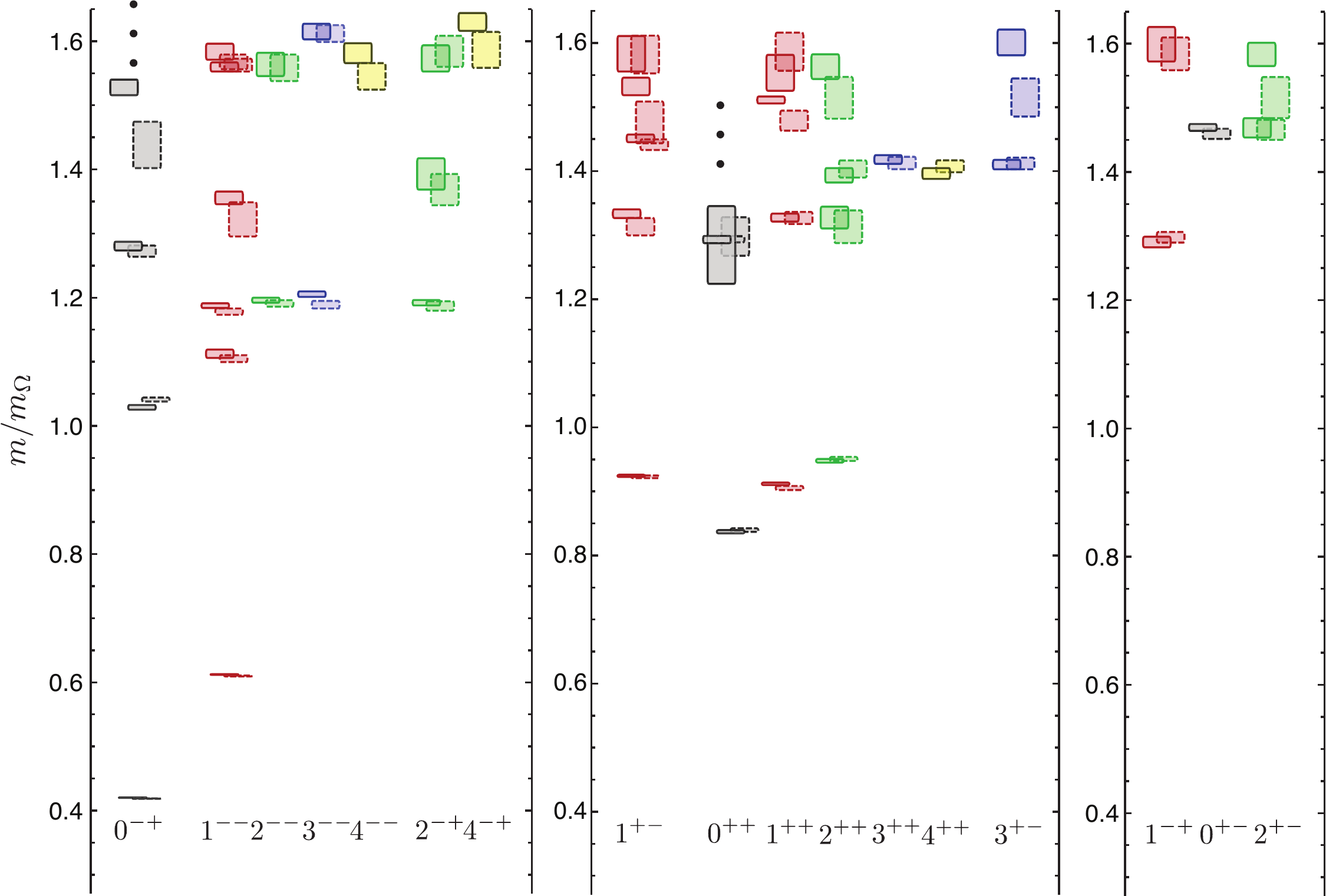}
\caption[]{\label{fig:lqcd_spect}(Color on line) The LQCD prediction for the spectrum of isovector mesons.
The quantum numbers are listed across the bottom, while the color denotes the spin. Solid (dashed) bordered boxes on a $2.0^3$($2.4^3)$ fm volume lattice, little volume dependence is observed. The
three columns at the far right are exotic-quantum numbers. The plot is taken from 
reference~\cite{Dudek:2010wm} .}
\end{figure*}

A significant LQCD calculation has recently been performed which predicts the entire 
spectrum of light-quark isovector mesons~\cite{Dudek:2009qf,Dudek:2010wm}. The
fully dynamical (unquenched) calculation is carried out with two flavors of the lightest quarks and 
a heavier third quark tuned to the strange quark mass. Calculations are performed on two lattice 
volumes and using four different masses for the lightest quarks---corresponding to pion masses of 
$700$, $520$, $440$ and $390$~MeV. In the heaviest case, the lightest quark masses are 
the same at the strange mass.  The computed spectrum of isovector states for this
heavy case is shown in Figure~\ref{fig:lqcd_spect} (where the mass is plotted as a ratio to 
the $\Omega$-baryon mass ($1.672$~GeV)). In the plot, the right-most columns correspond 
to the exotic $\pi_{1}$, $b_{0}$ and $b_{2}$ states. Interestingly, the $1^{-+}$ $\pi_{1}$ is the 
lightest, and both a ground state and what appears to be an excited state are predicted. 
The other two exotic-quantum-number states appear to be somewhat heavier than the 
$\pi_{1}$ with an excited state for the $b_{2}$ visible.

In addition to performing the calculation near the physical quark mass, there are a number 
of important innovations. First, the authors have found that the reduced rotational symmetry 
of a cubic lattice can be overcome on sufficiently fine lattices. They used meson operators of 
definite continuum spin \emph{subduced} into the irreducible representations of cubic rotations 
and observed very strong correlation between operators and the spin of the state. In this way they 
were able to make spin assignments from a single lattice spacing.  Second, the unprecedented 
size of the operator basis used in a variational calculation allowed the extraction of many excited 
states with confidence.

There were also phenomenological implications of these lattice results. A subset of
the meson operators feature the commutator of two gauge-covariant derivatives, 
equal to the field-strength tensor, which is non-zero only for non-trivial gluonic field 
configurations. Large overlap onto such operators was used to determine the degree 
to which gluonic excitations are important in the state, \emph{i.e.}, what one would call 
the \emph{hybrid} nature of the state. In particular, the exotic quantum number states 
all have large overlap with this type of operator, a likely indication of hybrid nature over, 
say, multiquark structure. In addition to the exotic-quantum number states, several
normal-quantum-number states also had large overlap with the non-trivial gluonic
field. In particular, states with $J^{PC}=1^{--}$, $2^{-+}$ with approximately the same mass 
as the lighter $1^{-+}$ state were noted.

\begin{figure}[h!]\centering
\includegraphics[width=0.45\textwidth]{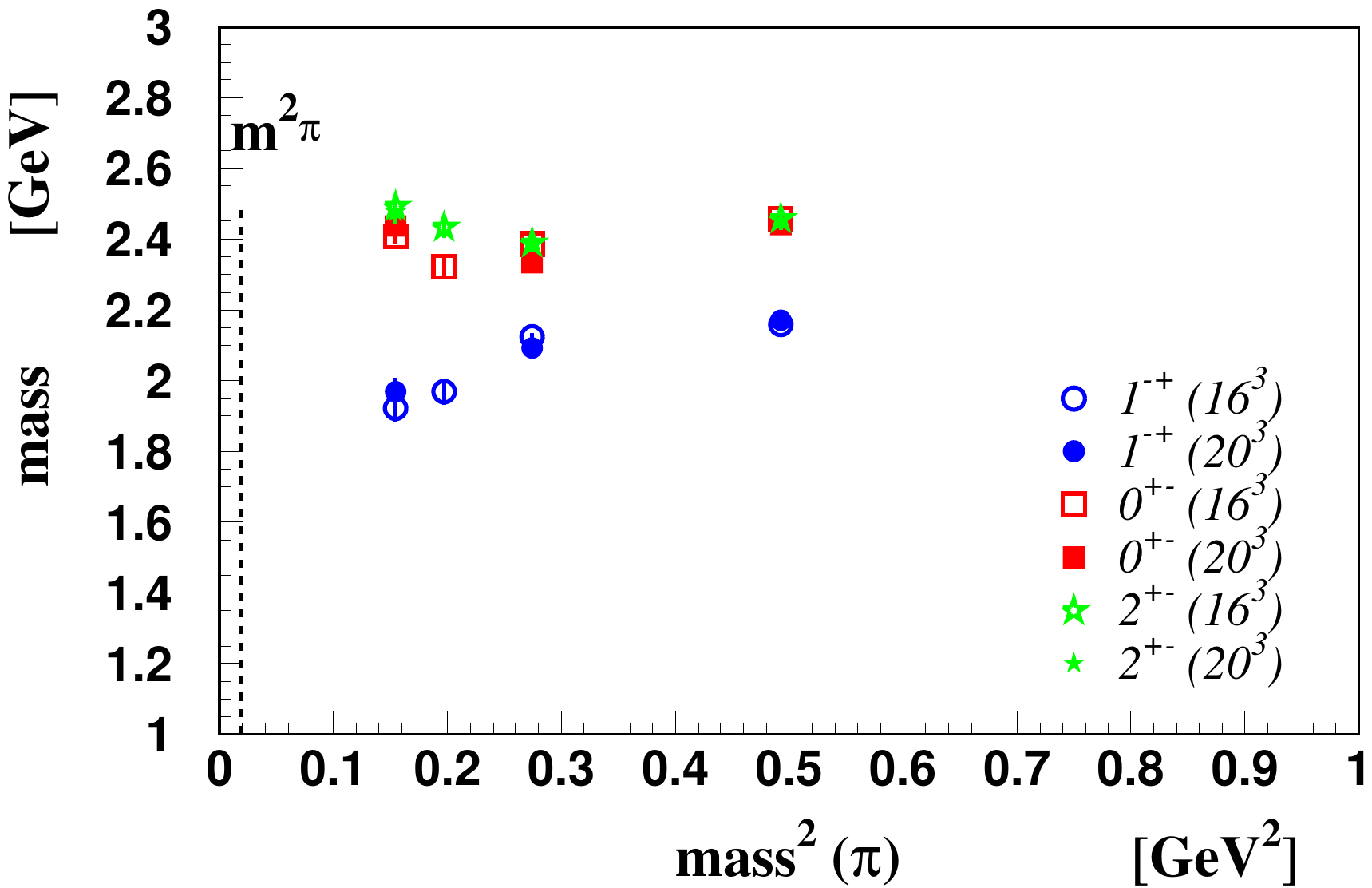}
\caption[]{\label{fig:mass_spectrum}(Color on line) The mass spectrum of the 
three exotic quantum number states~\cite{Dudek:2010wm}. The open figures are
for a $16^{3}$ spatial dimension lattice, while the solid are for a $20^{3}$ spatial
lattice. The (blue) circles are the mass of the $1^{-+}$ state, the (green) squares 
are the mass of the $0^{+-}$ state and the (red) stars are the $2^{+-}$ state.}
\end{figure}

In order to extract the masses of states, it is necessary to work at the physical pion mass. 
While work is currently underway to  extract a point at $m_{\pi}\approx 280$~MeV, this limit 
has not yet been reached. To attempt to extrapolate, one can plot the extracted state masses 
as a function of the pion mass squared, which acts as a proxy for the light quark mass (see 
Figure~\ref{fig:mass_spectrum}). While linearly extrapolating to the physical pion mass ignores 
constraints from chiral dynamics, it is probably safe to say that both the $\pi_1(1600)$ and the 
$\pi_{1}(2015)$ (as discussed below) could be consistent with the expected $1^{-+}$ mass. They 
are also consistent with the ground and first-excited $\pi_{1}$ state. It appears that the $b_{0}$ 
and $b_{2}$ masses will likely be several hundred MeV heavier than the lightest $\pi_{1}$. 

Lattice calculations have also been performed to look for other exotic quantum number states.
Bernard~\cite{bernard-97} included operators for a $0^{--}$ state, but found no evidence for
a state with these quantum numbers in their quenched calculation. 
Dudek \emph{et a.}~\cite{Dudek:2010wm}
looked for both $0^{--}$ and $3^{-+}$ states in their lattice data. They found some evidence 
for states with these quantum numbers, but the lightest masses were more than $2$~GeV 
above the mass of the $\rho$ meson.
 
These recent lattice calculations are extremely promising. They reaffirm that hybrid mesons form part of the low-energy QCD spectrum and that exotic quantum number states exist. They
also provide, for the first time, the possibility of assessing the gluonic content of
a calculated lattice state. Similar calculations are currently underway for the isoscalar
sector where preliminary results~\cite{dudek-priv} for the mass scale appear consistent with those shown
here in the isovector sector. These calculations will also extract the flavor mixing angle, an important quantity for phenomenology.

\subsection{\label{sec:decays}Decay Modes}
Currently, decays of hybrid mesons can only be calculated within models. Such models exist,
having been developed to compute the decays of normal mesons. A basic feature
of these is the so-called triplet-P-zero ($^{3}P_{0}$) model. In the
$^{3}P_{0}$ model, a meson decays by producing a $q\bar{q}$ pair with vacuum quantum 
numbers ($J^{PC}=0^{++}$).

A detailed study by Ackleh, Barnes and Swanson~\cite{Ackleh:1996} 
established that the $^{3}P_{0}$ amplitudes are dominant in
most light-quark meson decays. They also determined the parameters in decay models
by looking at the well known decays of mesons. This work was later extended to provide
predictions for the decay of all orbital and radial excitations of mesons lighter than 
$2.1$~GeV~\cite{Barnes:1997}. This tour-de-force in calculation has served as the 
backdrop against which most light-quark meson and hybrid candidates are compared.

The original calculations for the decays of hybrids in the flux-tube model were
carried out by Isgur~\cite{isgur-85b}.  Within their model, Close and Page~\cite{Close:1995}, confirmed the
results and expanded the calculations to include additional hybrids. Using improved
information about mesons and using simple harmonic oscillator (SHO) wave functions, they
were able to compute the decay width of hybrid mesons. They also provided arguments for 
the selection rule that hybrids prefer to decay to an $L=0$ and an $L=1$ meson. The 
suppression of a pair of $L=0$ mesons arises in the limit that the two mesons have the same 
inverse radius in the  Simple Harmonic Oscillator wave functions. Thus, these decays are not 
strictly forbidden, but are suppressed depending on how close the two inverse radii are. This 
led to the often-quoted predication for the decays of the $\pi_{1}$ hybrid given in 
equation~\ref{eq:decay_rates}.
\begin{eqnarray}
\nonumber
& \pi b_{1} : \pi f_{1} : \pi\rho : \eta\pi : \pi\eta^{\prime} &  \\ 
\nonumber
& = & \\
& 170 : 60 : 5-20 : 0-10 : 0-10 & 
\label{eq:decay_rates}
\end{eqnarray}

The current predictions for the widths of exotic-quantum-number hybrids are based 
on model calculations by Page \emph{et al.}~\cite{page-99} for which the results are
given in Table~\ref{tab:exotic-hybrid-widths}. They also computed decay rates for the 
hybrids with normal $q\bar{q}$ quantum numbers (results in Table~\ref{tab:non-exotic-hybrid-widths}). 
While a number of these states are expected to be broad (in particular, most of the $0^{+-}$ exotic nonet), 
states in both the $2^{+-}$ and the $1^{-+}$ nonets are expected to have much narrower widths. 
The expected decay modes involve daughters that in turn decay. Thus making the overall
reconstruction and analysis of these states much more complicated then simple 
two-pseudoscalar decays.
\begin{table}
\begin{tabular}{ccccc}\hline\hline
  &  &  &  & \\ 
Name & $\mathbf{J^{PC}}$ & \multicolumn{2}{c}{Total Width $MeV$} & 
Large Decays\\ 
         &                   & PSS & IKP & \\ \hline
$\pi_{1}$   & $1^{-+}$ &  $81-168$ & $117$ & 
$b_{1}\pi$, $\rho\pi$, $f_{1}\pi$, $a_{1}\eta$, \\
         & & & & $\eta(1295)\pi$, $K_{1}^{A}K$, $K_{1}^{B}K$ \\
$\eta_{1}$  & $1^{-+}$ &  $59-158$ & $107$ &
$a_{1}\pi$, $f_{1}\eta$, $\pi(1300)\pi$, \\
         & & & &  $K_{1}^{A}K$, $K_{1}^{B}K$ \\
$\eta^{\prime}_{1}$ & $1^{-+}$ &  $95-216$ & $172$ &
$K_{1}^{B}K$, $K_{1}^{A}K$, $K^{*}K$ \\ \hline
$b_{0}$     & $0^{+-}$ & $247-429$ & $665$ &
$\pi(1300)\pi$, $h_{1}\pi$ \\
$h_{0}$     & $0^{+-}$ & $59-262$  & $94$  &
$b_{1}\pi$, $h_{1}\eta$, $K(1460)K$ \\
$h^{\prime}_{0}$    & $0^{+-}$ & $259-490$ & $426$ &
$K(1460)K$, $K_{1}^{A}K$, $h_{1}\eta$ \\ \hline
$b_{2}$     & $2^{+-}$ &    $5-11$ & $248$ &
$a_{2}\pi$, $a_{1}\pi$, $h_{1}\pi$ \\
$h_{2}$     & $2^{+-}$ &    $4-12$ & $166$ &
$b_{1}\pi$, $\rho\pi$ \\
$h^{\prime}_{2}$    & $2^{+-}$ &    $5-18$ &  $79$ &
$K_{1}^{B}K$, $K_{1}^{A}K$, $K^{*}_{2}K$, $h_{1}\eta$ \\ 
\hline\hline 
\end{tabular}
\caption[Hybrid Widths]{\label{tab:exotic-hybrid-widths}
Exotic quantum number hybrid width and decay predictions from reference~\cite{page-99}. 
The column labeled PSS (Page, Swanson and Szczepaniak) is from their model, while the IKP 
(Isgur, Karl and Paton) is their calculation of the model
in reference~\cite{isgur-85b}. The variations in width for PSS come from different choices for
the masses of the hybrids. The $K_{1}^{A}$ represents the $K_{1}(1270)$ while the $K_{1}^{B}$ 
represents the $K_{1}(1400)$.}
\end{table}

For the non-exotic quantum numbers states, it will be even more difficult. They are likely 
to mix with nearby  normal $q\bar{q}$ states, complicating the expected decay pattern 
for both the hybrid and the normal mesons. However,  the decays in 
Table~\ref{tab:non-exotic-hybrid-widths} can be used as a guideline to help in identifying
these states. In searches for hybrid mesons,  the nonets with exotic quantum numbers 
provide the cleanest environment in which to search for these objects. 

Close and Thomas~\cite{Close:2009ii} reexamined this problem in terms of work on hadronic loops 
in the $c\bar{c}$ sector by Barnes and Swanson~\cite{Barnes:2007xu}. They conclude 
that in the limit where all mesons in a loop belong to a degenerate subset, vector hybrid
mesons remain orthogonal to the $q\bar{q}$ states ($J^{PC}=1^{--}$ $^{3}S_{1}$ and $^{3}D_{1}$) 
and mixing may be minimal. Thus, the search for hybrids with vector $q\bar{q}$ quantum numbers 
may not be as difficult as the other non-exotic quantum number hybrids.
\begin{table}[h!]\centering
\begin{tabular}{ccccc}\hline \hline
  &  &  &  & \\ 
Particle & $\mathbf{J^{PC}}$ & \multicolumn{2}{c}{Total Width $MeV$} & 
Large Decays\\ 
         &                   & PSS & IKP & \\ \hline
$\rho_{1}$      & $1^{--}$ & $70-121$ & $112$ & $a_{1}\pi$,$\omega\pi$, $\rho\pi$\\
$\omega_{1}$    & $1^{--}$ & $61-134$ & $60$  & $\rho\pi$, $\omega\eta$, 
$\rho(1450)\pi$\\
$\phi_{1}$      & $1^{--}$ & $95-155$& $120$ & $K_{1}^{B}K$, $K^{*}K$, 
$\phi\eta$ \\ \hline
$a_{1}$     & $1^{++}$ & $108-204$ & $269$ & $\rho(1450)\pi$, $\rho\pi$, 
$K^{*}K$\\
$h_{1}$     & $1^{++}$ & $43-130$  & $436$ & $K^{*}K$, $a_{1}\pi$    \\
$h^{\prime}_{1}$    & $1^{++}$ & $119-164$ & $219$ & $K^{*}(1410)K$,$K^{*}K$ \\ \hline
$\pi_{0}$       & $0^{-+}$ & $102-224$ & $132$ & $\rho\pi$,$f_{0}(1370)\pi$ \\
$\eta_{0}$      & $0^{-+}$ & $81-210$ & $196$ & $a_{0}(1450)\pi$, $K^{*}K$ \\
$\eta^{\prime}_{0}$     & $0^{-+}$ & $215-390$& $335$ & $K^{*}_{0}K$,$f_{0}(1370)\eta$,
$K^{*}K$ \\ \hline
$b_{1}$     & $1^{+-}$ & $177-338$ & $384$ & $\omega(1420)\pi$,$K^{*}K$ \\
$h_{1}$     & $1^{+-}$ & $305-529$ & $632$ & $\rho(1450)\pi$, $\rho\pi$, 
$K^{*}K$ \\
$h^{\prime}_{1}$    & $1^{+-}$ & $301-373$ & $443$ & $K^{*}(1410)K$, $\phi\eta$, 
$K^{*}K$ \\
$\pi_{2}$   & $2^{-+}$ & $27-63$ & $59$ &  $\rho\pi$,$f_{2}\pi$ \\
$\eta_{2}$  & $2^{-+}$ & $27-58$ & $69$ & $a_{2}\pi$ \\
$\eta^{\prime}_{2}$ & $2^{-+}$ & $38-91$  & $69$  & $K^{*}_{2}K$, $K^{*}K$ \\ 
\hline\hline
\end{tabular}
\caption[Hybrid Widths]{\label{tab:non-exotic-hybrid-widths}
Non-exotic quantum number hybrid width and decay predictions from reference~\cite{page-99}. 
The column labeled PSS (Page, Swanson and Szczepaniak) is from their model, 
while the IKP (Isgur, Karl and Paton) is their calculation of the model
in reference~\cite{isgur-85b}. The variations in width for PSS come from different choices for
the masses of the hybrids. The $K_{1}^{A}$ represents the $K_{1}(1270)$ while the $K_{1}^{B}$ 
represents the $K_{1}(1400)$.}
\end{table}

Almost all models of hybrid mesons predict that they will not decay to identical pairs of 
mesons. Many also predict that decays to pairs of $L=0$ mesons will be suppressed, leading
to decays of an $(L=0)(L=1)$ pair as the favored hybrid decay mode. 
Page~\cite{Page:1996rj} undertook a study of these models of hybrid decay that included 
``TE hybrids'' ( with a transverse electric constituent gluon) in the bag model  as well as 
``adiabatic hybrids'' in the flux-tube model (hybrids in the limit where quarks move  slowly 
with respect to the gluonic degrees of freedom). In such cases, the decays to pairs of 
orbital angular momentum $L=0$ (S–wave) mesons were found to vanish.  In both cases, it 
had been noted that this was true when the quark and the antiquark in the hybrid's daughters 
have identical constituent masses with the same $S$-wave spatial wave functions, and the 
quarks are non-relativistic. In order to understand  this, Page looked for an underlying 
symmetry that could be responsible for this.
\begin{figure}[h!]\centering
\begin{tabular}{cc}
\includegraphics[width=0.25\textwidth]{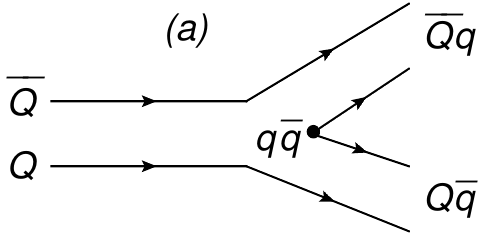} &
\includegraphics[width=0.15\textwidth]{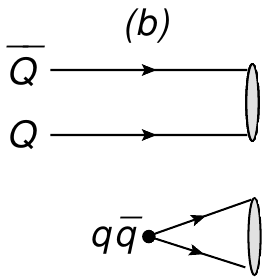}
\end{tabular}
\caption[]{\label{fig:suppress} (a) shows a connected decay diagram where the
decay can be suppressed. (b) is an example of a disconnected diagram where 
the decay is not suppressed..}
\end{figure}
He found that symmetrization of connected decay diagrams (see Figure~\ref{fig:suppress}(a) )
where the daughters are identical except for flavor and spin vanish when equation~\ref{eq:page}
is satisfied.
\begin{eqnarray}
\label{eq:page}
C^{0}_{A}P_{A} & = & (-1)^{(S_{A}+S_{q\bar{q}}+1)}
\end{eqnarray}
For meson $A$ decaying to daughters $B$ and $C$, $C^{0}_{A}$ is the C-parity of the neutral 
isospin member of the decaying meson $A$, $P_{A}$ is its parity and $S_{A}$ is its intrinsic 
spin. $S_{q\bar{q}}$ is the total spin of the created pair. In the non-relativistic limit, $S_{q\bar{q}}=1$. 
For non-connected diagrams (Figure~\ref{fig:suppress}(b)), he found no such general rules, 
so the vanishing of the decays occur to the extent that the non-connected diagrams are not 
important (OZI suppression).

As an example of this, consider $A$ to be the $\pi_{1}$ hybrid. It has $C^{0}_{A}=+1$, $P_{A}=-1$ 
and $S_{A}=+1$, thus the left-hand side of equation~\ref{eq:page} is $-1$.  The right-hand side 
is $(-1)^{3}=-1$. The decay to pairs of mesons with the same internal angular momentum
is suppressed to the extent that the disconnected diagram in Figure~\ref{fig:suppress}(b) is
not important. In a later study, Close and Dudek~\cite{Close:2003af} found that some of these decays
could be large because the $\pi$ and $\rho$ wave functions were not the same.

While it has been historically difficult to compute decays on the lattice, a first study
of the decay of the $\pi_{1}$ hybrid has been carried out by McNeile~\cite{McNeile:2002,McNeile:2006bz}.
 In order to do this, they used
a technique where they put a given decay channel at roughly the same energy as the 
decaying state. Thus,  the  decay is just allowed and conserves energy in a two-point 
function. In this way, they are able to extract the ratio of the decay width over the
decay momentum, and find 
\begin{eqnarray*}
\Gamma (\pi_{1}\rightarrow b_{1}\pi)/k & = & 0.66\pm 0.20 \\
\Gamma (\pi_{1}\rightarrow f_{1}\pi)/k & = & 0.15\pm 0.10  
\end{eqnarray*} 
which they note corresponds to a total decay width larger than $0.4$~GeV for
the $\pi_{1}$. As a check of their procedure, they carry out a similar calculation for 
$b_{1}\rightarrow \omega \pi$ where they obtain $\Gamma/k\sim 0.8$, which 
leads to $\Gamma(b_{1}\rightarrow\omega\pi)\sim 0.22$~GeV. This is about a
factor of $1.6$ larger than the experimental width. 

Burns and Close~\cite{Burns:2006} examined these lattice decay results and made
comparisons to what had been found in flux-tube model calculations. In Table~\ref{tab:ft_lt_dcy}
are shown their comparison between flux-tube calculations for the width of the $\pi_{1}$ and
the decay width from the lattice. They note that in the work of McNeile~\cite{McNeile:2006bz}, an 
assumption was made
that $\Gamma/k$ does not vary with the quark mass, and the resulting linear extrapolation
leads to the large width in Table~\ref{tab:ft_lt_dcy}. They argue that the flux-tube model has
been tested over a large range of $k$, where it accurately predicts the decays of mesons and
baryons. Quoting them, ``The successful phenomenology of this and a wide range of other 
conventional meson decays relies on momentum-dependent form factors arising from the 
overlap of hadron wave functions. The need for such form factors is rather general, empirically 
supported as exclusive hadron decay widths do not show unrestricted growth with phase space.''
Based on this, they carried out a comparison of the transition amplitudes computed for $k=0$
(the lattice case). They found excellent agreement between the lattice and the flux-tube calculations.
Thus, their concern that the extrapolation may be overestimating decay widths may be valid. 
\begin{table}[h!]\centering
\begin{tabular}{rccc}\hline\hline
 & IKP & IKP & Lattice \\
 & \cite{isgur-85b} & \cite{Close:1995} & \cite{McNeile:2006bz} \\
 & $1.9$~GeV & $2.0$~GeV & $2.0$~GeV \\ \hline
$\Gamma(\pi_{1}\rightarrow b_{1}\pi)_{S}$ & $100$ & $70$ & $400\pm 120$ \\
$\Gamma(\pi_{1}\rightarrow b_{1}\pi)_{D}$ & $30$   & $30$ &  \\
$\Gamma(\pi_{1}\rightarrow f_{1}\pi)_{S}$ & $30$    & $20$ & $90\pm 60$ \\
$\Gamma(\pi_{1}\rightarrow f_{1}\pi)_{D}$ & $20$    & $25$ &  \\
\hline\hline
\end{tabular}
\caption[]{\label{tab:ft_lt_dcy} Decay widths as computed in the flux-tube model (IKP) 
compared to the lattice calculations. (Table reproduced from reference~\cite{Burns:2006}.)}
\end{table}

While the model calculations provide a good guide in looking for hybrids, there are 
often symmetries that can suppress or enhance certain decays. Chung and 
Klempt~\cite{Chung:2002fz} noted one of these for decays of a $J^{PC}=1^{-+}$ state
into $\eta\pi$  where the $\eta$ and $\pi$ have relative angular momentum of
$L=1$. In particular, in the limit where the $\eta$ is an SU(3) octet, the $\eta\pi$ in
a $p$-wave must be in an antisymmetric wave function. In order to couple this to 
an octet (hybrid) meson, the hybrid must also be antisymmetric. This implies it 
must be a member of the $8_{2}$ octet. However, the SU(3) Clebsch-Gordan coefficient for 
$8_{2}\rightarrow \eta_{8}\pi$ is zero. Thus, the decay is forbidden. 

However, by similar
arguments, they showed that it can couple to the  $10\oplus\overline{10}$ representation
of SU(3). A representation that contains multiquark ($qq\bar{q}\bar{q}$) objects
(see Section~\ref{sec:multiquark}). Similarly, for the  singlet ($\eta^{\prime}$) in a $p$-wave,  
the coupling to an octet is not suppressed. 

To the extent that the $\eta$ is octet and the $\eta^{\prime}$ is singlet, a $1^{-+}$ state
that decays to $\eta\pi$ and not $\eta^{\prime}\pi$ cannot be a hybrid, while one that
decays to $\eta^{\prime}\pi$ and not $\eta\pi$ is a candidate for a $1^{-+}$ hybrid meson.
Our current understanding of the pseudoscalar mixing angle is that it is between $-10^{\circ}$ 
and $-20^{\circ}$~\cite{amsler08}, thus the assumption on the nature of the $\eta$ and
$\eta^{\prime}$ is not
far off. However, as far as we know, the pseudoscalar mesons are the only nonet that
is close to pure SU(3) states, all others tend to be close to ideal mixing. A case where 
the higher mass state is nearly pure $s\bar{s}$. Thus, this suppression would not be
expected for decays to higher mass nonets.

\subsection{\label{sec:multiquark}Multiquark states}
As noted in Section~\ref{sec:intro}, exotic quantum numbers can arise from other
quark-gluon systems as well. While it is possible for glueballs to have exotic quantum 
numbers, the masses are expected to be above $3$~GeV~\cite{Morningstar:1999rf}.
Another configuration are multiquark states ($qq\bar{q}\bar{q}$) consisting of two quarks
and two antiquarks. A short review of this topic can be found in Ref.~\cite{Roy:2003hk}, and
a nice description of how these states are built in the quark model can be found in 
Ref.~\cite{Close-Quark-Model}.

Following Ref.~\cite{Jaffe:1977cv}, the SU(3) multiplets of these states can be obtained by 
considering $qq$ and $\bar{q}\bar{q}$ combinations. The former can transform as either 
$\overline{3}$ or $6$ under SU(3), while the latter can transform as $3$ and $\overline{6}$. Thus, 
multiplets can be built up as
\begin{eqnarray*}\begin{array}{ccccc}
\overline{3} \otimes 3 & = & 1 \oplus 8 &  = & 9 \\
6 \otimes \overline{6} &  = & 1 \oplus 8 \oplus 27 & = & 36 \\
6 \otimes 3 \oplus \overline{3} \otimes \overline{6} & = & 8 \oplus 10 \oplus 8 \oplus \overline{10} & = & 18 \oplus \overline{18} \end{array} 
\end{eqnarray*}
The $J^{P}$ of these multiquark states can be obtained by initially combining all the quarks in
an S-wave. This yields $J^{P}$ values of $0^{+}$, $1^{+}$ and $2^{+}$, which can be combined 
with the fact that the overall wave functions must be antisymmetric to associate SU(3) multiplets 
with $J^{P}$. 
\begin{eqnarray*}
J^{P} = 2^{+} & :  & 9,36 \\
J^{P} = 1^{+} & :  & 9, 18, 18, \overline{18}, \overline{18}, 36 \\
J^{P} = 0^{+} & :  & 9, 9, 36, 36 \\
\end{eqnarray*}
Jaffe considered these multiquark states in terms of the bag model~\cite{Jaffe:1977cv,Jaffe:1976ih},
where he found a nonet of $J^{P}=0^{+}$ states to be the lightest with a mass around $1$~GeV. 
This \emph{cryptoexotic} nonet is interesting in that the $\rho$- and $\omega$-like states have 
an $s\bar{s}$ pair combined with the lighter quarks.
\begin{eqnarray*}
\omega   & \,\,\,\, & \frac{1}{\sqrt{2}} \left( u\bar{u} + d\bar{d} \right) (s\bar{s}) \\
\rho^{+}   & \,\,\,\, &  u\bar{d} (s\bar{s})\\
\rho^{0}    & \,\,\,\, &  \frac{1}{\sqrt{2}} \left( u\bar{u} - d\bar{d} \right) (s\bar{s})\\
\rho^{-}    & \,\,\,\, &  d\bar{u} (s\bar{s})
\end{eqnarray*} 
The $K$-like states have a single strange quark, 
\begin{eqnarray*}
K^{+}           & \,\,\,\, & u\bar{s}  d\bar{d} \\
K^{0}            & \,\,\,\, & d\bar{s} u\bar{u} \\
\bar{K}^{0}    & \,\,\,\, & s\bar{u} d\bar{d}  \\
K^{-}            & \,\,\,\, & s\bar{d} u\bar{u} 
\end{eqnarray*} 
while the $\phi$-like state has no strange quarks, 
\begin{eqnarray*}
\phi           & \,\,\,\, & u\bar{u}  d\bar{d} \, .
\end{eqnarray*}
This yields the so-called inverted nonet, where the mass-hierarchy is reversed relative
to the $q\bar{q}$ states. This nonet is often associated with the 
low-mass states $f_{0}(600)$ ($\sigma$), $K^{*}_{0}(800)$ ($\kappa$), $a_{0}(980)$ and
the $f_{0}(980)$. Jaffe also noted that whenever the expected mass of a multiquark
state was above that of a simple meson-meson threshold to which the state could couple, 
the decays would be ``super-allowed'', and the width of the state would be very large. 
Because of these super-allowed decays, Jaffe~\cite{Jaffe:79} noted that the states would 
not exist.

Orbital excitations of the multiquark systems were examined in reference~\cite{Aerts:1979hn}.
Additional symmetrization rules beyond the simple $q\bar{q}$ system apply for these, but
they found that the addition of one unit of angular momentum could produce both 
$J^{PC}=1^{-+}$ and $0^{--}$ states as members of an $18\oplus\overline{18}$ SU(3) multiplet
with masses around $1.7$~GeV. There are two isovector states in an $18$, one as part of 
an octet and the second as part of a decuplet. The multiquark representation can be 
represented In a meson-meson-like by recoupling the colors and spins to the new basis. 
Doing this,  the two isovector states look like a $\pi$ combined with either and $\eta$ or 
an $\eta^{\prime}$. While the mixing between the $\eta$ and $\eta^{\prime}$ components 
is not known, it is likely that both states would have some hidden $s\bar{s}$ component.

General and colleagues~\cite{General:2007bk} looked at multiquark states in the 
framework of molecular resonances using their coulomb gauge formalism. In this 
framework, they computed the spectrum of the lightest states and find several states
with masses below $1.5$~GeV. In the isovector sector, they find the lightest state to
be a $J^{PC}=1^{-+}$ state ($m=1.32$~GeV), with a somewhat heavier $0^{--}$ state
($m=1.36$~GeV), and then a second $1^{-+}$ state ($m=1.42$~GeV). In the isoscalar 
channel, they find a single $0^{--}$ state and in the isotensor (isospin two) channel, they 
predict an additional $0^{--}$ state. Between $1.5$ and $2$~GeV, they predict two additional 
$1^{-+}$ states in each of the three isospin channels.  

QSSR techniques have also been used to look for both isovector~\cite{Chen:2008qw} and
isoscalar~\cite{Chen:2008ne} $J^{PC}=1^{-+}$ multiquark states. As with the earlier work,
they find that the exotic-quantum number multiquark states are in the  
$(\overline{3}\otimes\overline{6})\oplus(3\otimes 6)$ flavor representations. In their
calculations, the decuplet $\pi_{1}$ state (with no $s\bar{s}$ pair) has a mass of about 
$1.6$~GeV, while the octet $\pi_{1}$ state (with $s\bar{s}$) has a mass of about $2$~GeV.
For these states, they suggest decays of the form $J^{P}=0^{+},J^{P}=1^{-}$ ($f_{0}\rho$),  
$J^{P}=1^{+},J^{P}=0^{-}$ ($b_{1}\pi$) and $J^{P}=1^{-},J^{P}=1^{+}$ ($\omega b_{1}$). 
For the isoscalar masses, both the octet and decuplet member contain an $s\bar{s}$ pair. They
find a single state with a mass between $1.8$ and $2.1$~GeV. For decays, their calculations 
favor decays of the form $K\overline{K}$, $\eta\eta$, $\eta\eta^{\prime}$ and 
$\eta^{\prime}\eta^{\prime}$. They also\begin{eqnarray*}
Z_{eq} & = & Z_{L} + Z_{C} \\
Z_{eq} & = & j\omega L + \frac{1}{j\omega C} \\
Z_{eq} & = & \frac{1 - \omega^{2}LC}{j\omega C} 
\end{eqnarray*}
 suggest several decays that are forbidden by isospin
conservation.

Lattice calculations for multiquark states are somewhat sparse, largely due to the challenge
of the number of quarks. Studies have been made to try to determine if the low-mass scalars
have multiquark nature. A calculation in the quenched approximation was made with pion
mass as small as $180$~MeV identified the $f_{0}(600)$ as a multiquark state~\cite{Mathur:2006bs}.
A later quenched calculation with heavier pion masses ($344$-$576$~MeV) found no indication of 
the $f_{0}(600)$~\cite{Prelovsek:2008rf}, but the authors note that their pion mass is too heavy for 
this to be conclusive. A recent dynamical calculation~\cite{Prelovsek:2010kg} with somewhat 
heavier pion mass shows good agreement with Ref.~\cite{Mathur:2006bs}, and while the authors
could not exclude the states are lattice artifacts, their results suggest that the $f_{0}(600)$ and
$K^{*}_{0}(800)$ have a multiquark nature. Finally, a recent dynamical calculation of the entire
isovector meson spectrum shows no multiquark states~\cite{Dudek:2010wm}. However, the authors
note that the correct operators were probably not included in their analysis, so the fact that these
states are missing from their analysis should not be taken as conclusive. Other lattice calculations 
explicitly looking for exotic-quantum-number multiquark states do not appear to have been
performed.

If exotic-quantum number multiquark states exist, the favoured quantum numbers are 
$1^{-+}$ and $0^{--}$. The latter being a $J^{PC}$ not predicted for hybrid mesons. There
may also be hidden $s\bar{s}$ components in the multiquark multiplets that would distort their
mass hierarchy relative to hybrid nonets. However, for most of these multiquark states, their
decays will be super-allowed. In their recent review, Klempt and Zaitsev~\cite{klempt-07} argue 
that $(qq)(\bar{q}\bar{q})$ systems will not bind without additional $q\bar{q}$ forces, and feel
that it is unlikely that these multiquark states exist. In reviewing the information on these states,
we concur with their assessment for the exotic-quantum-number states.

\section{\label{sec:results}Experimental Results}
\subsection{\label{sec:production}Production processes}
\begin{figure}[h!]\centering
\includegraphics[width=0.25\textwidth]{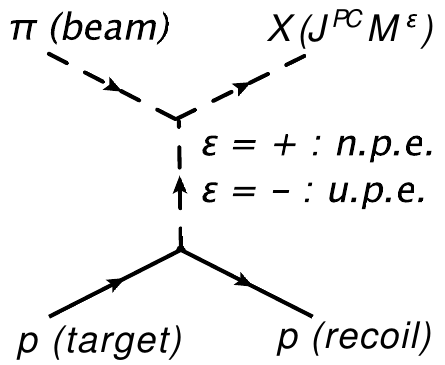}
\caption[]{\label{fig:t-channel}The diffractive production process showing an incident 
pion ($\pi$ beam) incident on a proton ($p$ target) where the exchange has z-component
on angular momentum $M$ and reflectivity $\epsilon$. The final state consists
of a proton ($p$ recoil) and a state $X$ of given  $J^{PC}$ produced by an exchange 
$M^{\epsilon}$.  For positive reflectivity, the $t$-channel is a natural parity exchange (n.p.e.),
while for negative reflectivity, it is unnatural parity exchange (u.p.e.). (This diagram was 
produced using the JaxoDraw package~\cite{jaxodraw}.)}
\end{figure}

Data on exotic-quantum-number mesons have come from both diffractive production using
incident pion beams and from antiproton annihilation on protons and neutrons. Diffractive 
production is schematically shown in Figure~\ref{fig:t-channel}. A pion beam is incident 
on a proton (or nuclear) target, which recoils after exchanging something in the $t$-channel.
The process can be written down in the reflectivity basis~\cite{Chung:1974fq} in which the 
production factorizes into two non-interfering amplitudes---positive reflectivity ($\epsilon=+$)
and negative reflectivity ($\epsilon=-$). The absolute value of the spin projection along the 
$z$-axis is $M$, and is taken to be either $0$ or $1$ (it is usually assumed that contributions from
$M$ larger than $1$ are small and can be ignored~\cite{chung-99}). It can be shown in this process that 
naturality of the exchanged particle can be determined by $\epsilon$. Natural parity 
exchange (n.p.e.) corresponds to $J^{P}$s of $0^{+}$, $1^{-}$, $2^{+}$, $\cdots$, while unnatural 
parity exchange (u.p.e.) corresponds to $J^{P}$ of $0^{-}$, $1^{+}$, $2^{-}$, $\cdots$.

For a state which is observed in more than one decay mode, one would expect that the production
mechanism ($M^{\epsilon}$) would be the same for all decay modes. If not, this could be indicative 
of more than one state being observed, or possible problems in the analysis that are not under
control. 

In antiproton-nucleon annihilation, there are a number of differences between various annihilation
processes. For the case of $\bar{p}p$, the initial state is a mixture of isospin $I=0$ and $I=1$. For
$\bar{p}n$ annihilation, the initial state is pure $I=1$. For annihilation at rest on protons, the 
initial state is dominated by atomic S-waves. In particular, $^{1}S_{0}$ and $^{3}S_{1}$ atomic states, 
which have $J^{PC}=0^{-+}$ and $1^{--}$ respectively (with a small admixture of $P$ states). For 
annihilation in flight, the number of initial states is much larger and it may no longer make 
sense to try and parametrize the initial system in terms of atomic states. 

The combination of initial isospin and final state particles may lead to additional selection 
rules that restrict the allowed initial states. In the  case of $\bar{p}p\rightarrow \eta\pi^{0}\pi^{0}$, 
the annihilation is dominated by $^{1}S_{0}$ initial states ($J^{PC}=0^{-+}$). For the case of 
$\bar{p}n\rightarrow \eta\pi^{0}\pi^{-}$, quantum numbers restrict this annihilation to occur from 
the $^{3}S_{1}$ initial states ($J^{PC}=1^{--}$). In addition, the neutron is bound in deuterium, where
the Fermi motion introduces substantial p-wave annihilation. Thus, one may see quite different 
final states from the two apparently similar reactions.

\begin{figure*}[t]\centering
\includegraphics[width=0.75\textwidth]{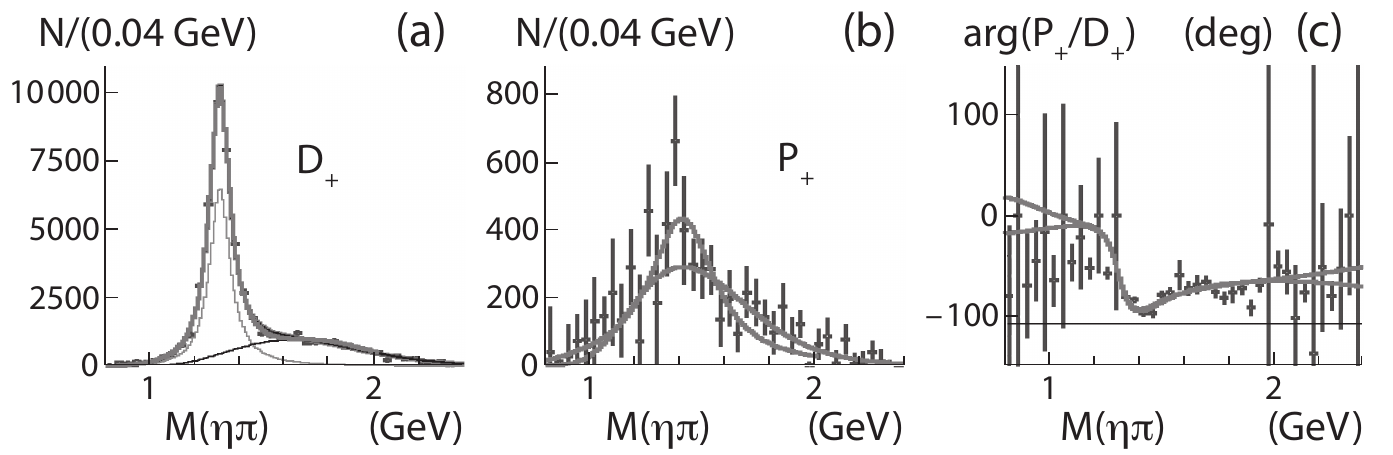}
\caption[]{\label{fig:ves_etapi}The results of a partial-wave analysis of the $\eta\pi^{-}$
final state from VES. (a) shows the intensity in the $2^{++}$ partial wave, (b) shows the 
intensity in the $1^{-+}$ partial wave and (c) shows the relative phase between the waves.
(Figure reproduced from reference~\cite{Amelin:2005ry}.)}
\end{figure*}
\subsection{\label{sec:p1_1400}The $\pi_{1}(1400)$}
The first reported observation of an exotic quantum number state came from the GAMS
experiment which used a $40$~GeV/c $\pi^{-}$ to study the reaction 
$\pi^{-}p\rightarrow p \eta \pi^{-}$. They reported a $J^{PC}=1^{-+}$ state in the 
$\eta\pi^{-}$ system which they called the $M(1405)$~\cite{alde-88}. The $M(1405)$ had
a mass of $1.405\pm 0.020$~GeV and a width of $0.18\pm 0.02$~GeV.
Interestingly, an earlier search in the $\eta\pi^{0}$ channel found no evidence of an
exotic state~\cite{appel-81}.  At KEK, results were reported on studies using a $6.3$~GeV/c
$\pi^{-}$ beam where they observed a $1^{-+}$ state in the $\eta \pi^{-}$ system with a
mass of $1.3431\pm 0.0046$~GeV and a width of $0.1432\pm 0.0125$~GeV~\cite{aoyagi-93}.
However, there was concern that this may have been leakage from the $a_{2}(1320)$.

The VES collaboration reported intensity in the $1^{-+}$ $\eta\pi^{-}$ wave as well as
rapid phase motion between the $a_{2}$ and the exotic wave~\cite{Beladidze:93}
(see Figure~\ref{fig:ves_etapi}). 
The exotic wave was present in the $M^{\epsilon}=1^{+}$ (natural parity) exchange, but 
not in the $0^{-}$ and $1^{-}$ (unnatural parity) exchange. They could fit the observed 
$J^{PC}=1^{-+}$ intensity and the phase motion with respect to the $a_{2}(1320)$ using
a Breit-Wigner distribution (mass of $1.316\pm 0.012$~GeV and width of $0.287\pm 0.025$~GeV).
However, they stopped short of claiming an exotic resonance, as they could not unambiguously 
establish the nature of the exotic wave~\cite{Dorofeev:02}. In a later analysis of the $\eta\pi^{0}$ 
system, they claim that the peak near $1.4$~GeV can be understood without requiring an exotic 
quantum number meson~\cite{Amelin:2005ry}.

\begin{figure}[h]\centering
\includegraphics[width=0.5\textwidth]{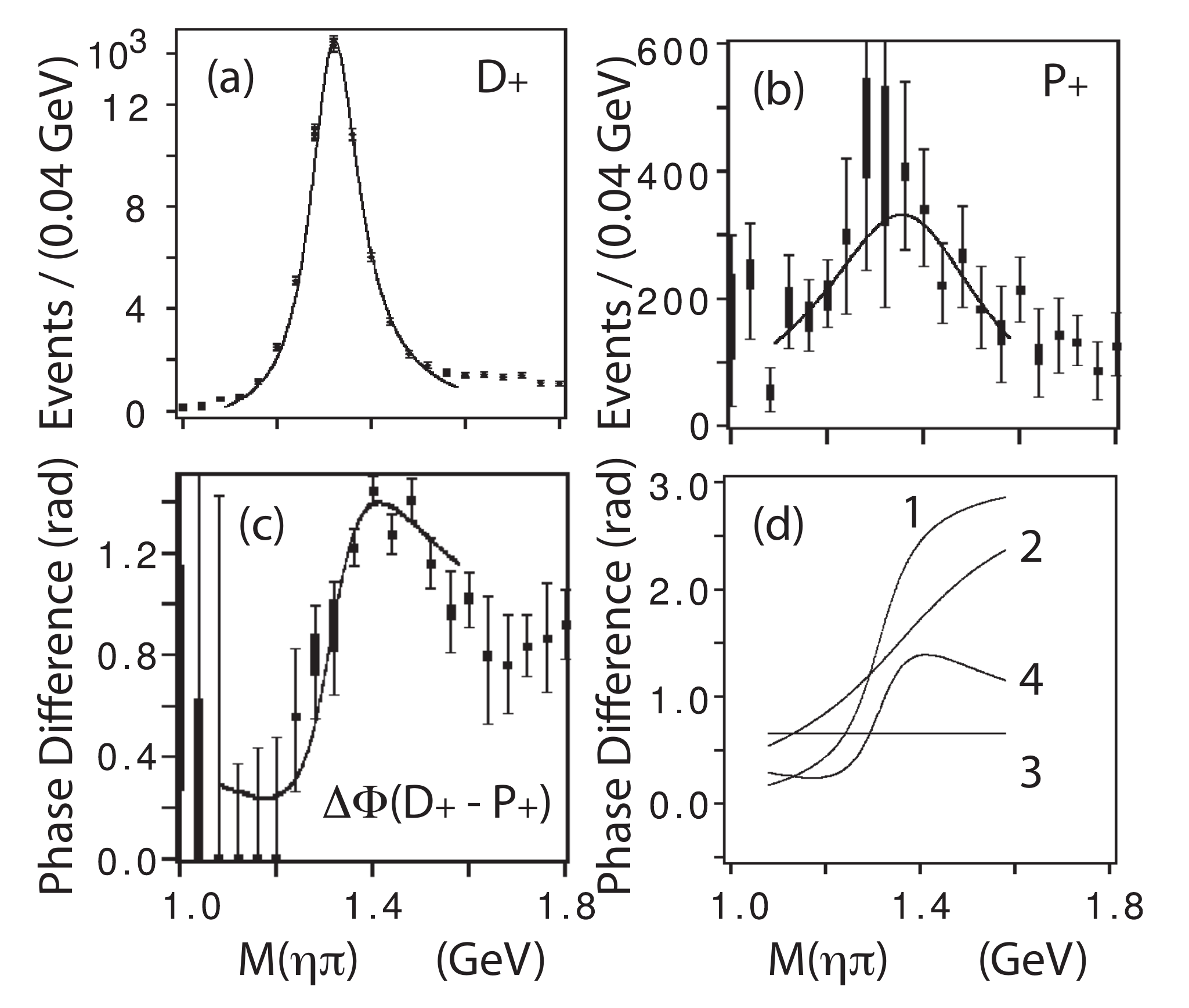}
\caption[]{\label{fig:e852_p1_1400}The $\pi_{1}(1400)$ as observed in the E852
experiment~\cite{Thompson:1997bs}. (a) shows the intensity of the $J^{PC}=2^{++}$ partial
wave as a function of $\eta\pi$ mass. The strong signal is the $a_{2}(1320)$. (b) shows
the intensity of the $1^{-+}$ wave as a function of mass, while (c) shows the phase 
difference between the $2^{++}$ and $1^{-+}$ partial waves. In (d) are shown the 
phases associated with (1) the $a_{2}(1320)$, (2) the $\pi_{1}(1400)$, (3) the assumed 
flat background phase and (4), the difference between the (1) and (2).
(This figure is reproduced from reference~\cite{Thompson:1997bs}.)}
\end{figure}

The E852 collaboration used $18$~GeV/c $\pi^{-}$ beams to study the reaction 
$\pi^{-}p\rightarrow p \eta \pi^{-}$. They reported the observation of a $1^{-+}$ 
state in the $\eta\pi^{-}$ system~\cite{Thompson:1997bs}. E852 found this state 
only produced in natural parity exchange ($M^{\epsilon}=1^{+}$). They measured a mass
of $1.37\pm 0.016^{+0.050}_{-0.030}$~GeV and a width of $0.385\pm 0.040^{+0.065}_{-0.105}$~GeV.
While the observed exotic signal was only a few percent of the dominant $a_{2}(1320)$ strength, 
they noted that its interference with the $a_{2}$ provided clear evidence of this state.
When their intensity and phase-difference plots were compared with those from
VES~\cite{Beladidze:93}, they were identical. These plots (from E852) are reproduced in
Figure~\ref{fig:e852_p1_1400}. 

Due to disagreements over the interpretation of the $1^{-+}$ signal, the E852 collaboration
split into two groups. The majority of the collaboration published the resonance interpretation,
$\pi_{1}(1400)$~\cite{Thompson:1997bs}, while a subset of the collaboration did not sign
the paper. As this latter group, centered at Indiana University, continued to analyze data 
collected by E852, we will refer to their publications as E852-IU to try an carefully distinguish 
the work of the two groups.

\begin{figure}[h!]\centering
\includegraphics[width=0.5\textwidth]{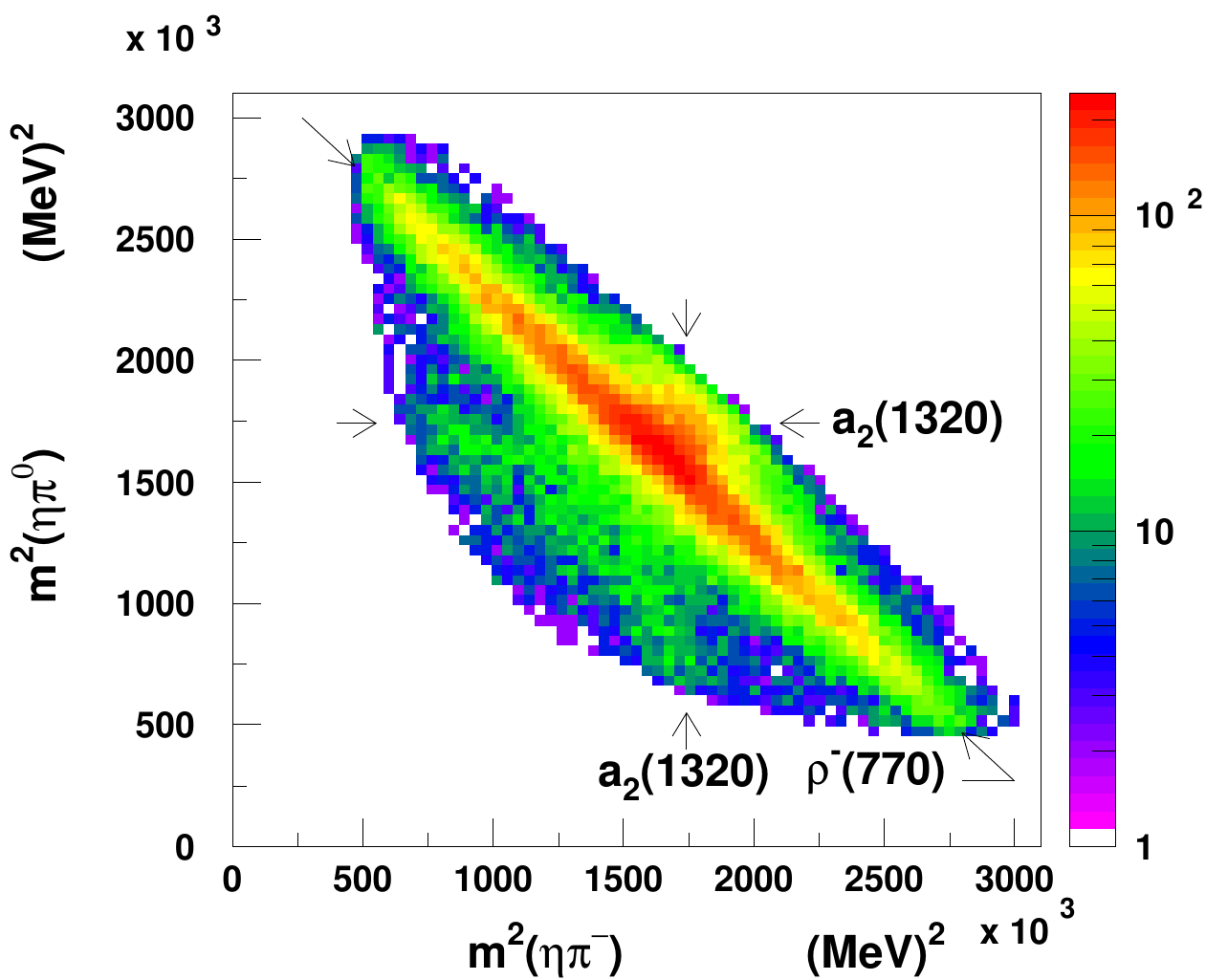}
\caption[]{\label{fig:cbar_p1_1400} (Color on line.) 
The Dalitz plot of $m^{2}(\eta\pi^{0})$ versus $m^{2}(\eta\pi^{-})$
for the reaction $\bar{p}n\rightarrow \eta\pi^{-}\pi^{0}$ from the Crystal Barrel 
experiment~\cite{cbar98}. The bands for the $a_{2}(1320)$ are clearly seen in both $\eta\pi^{0}$ and
$\eta\pi^{-}$, while the $\rho(770)$ is seen in the $\pi^{0}\pi^{-}$ invariant mass.}
\end{figure}

\begin{figure}[h!]\centering
\begin{tabular}{c}
\includegraphics[width=0.5\textwidth]{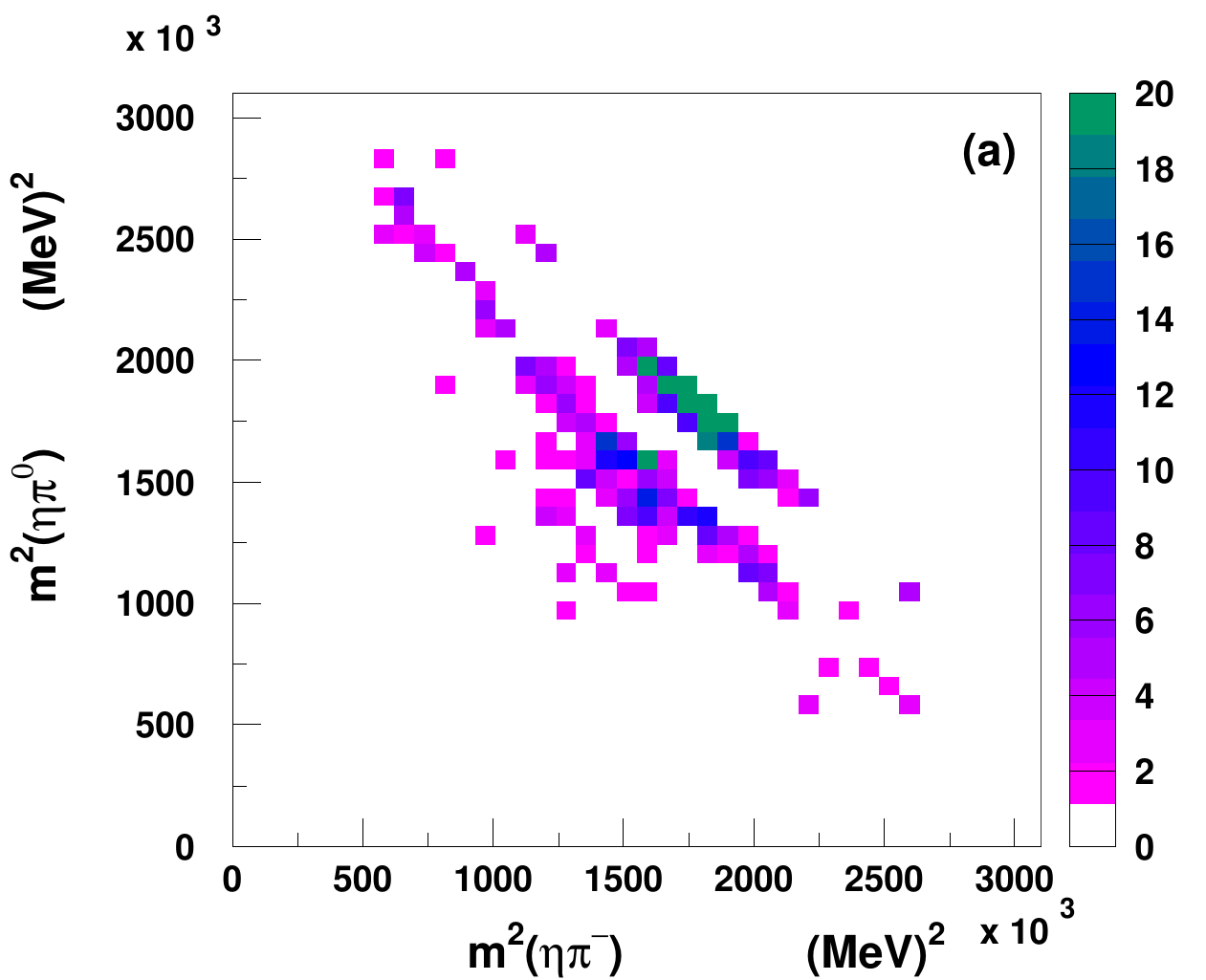} \\
\includegraphics[width=0.5\textwidth]{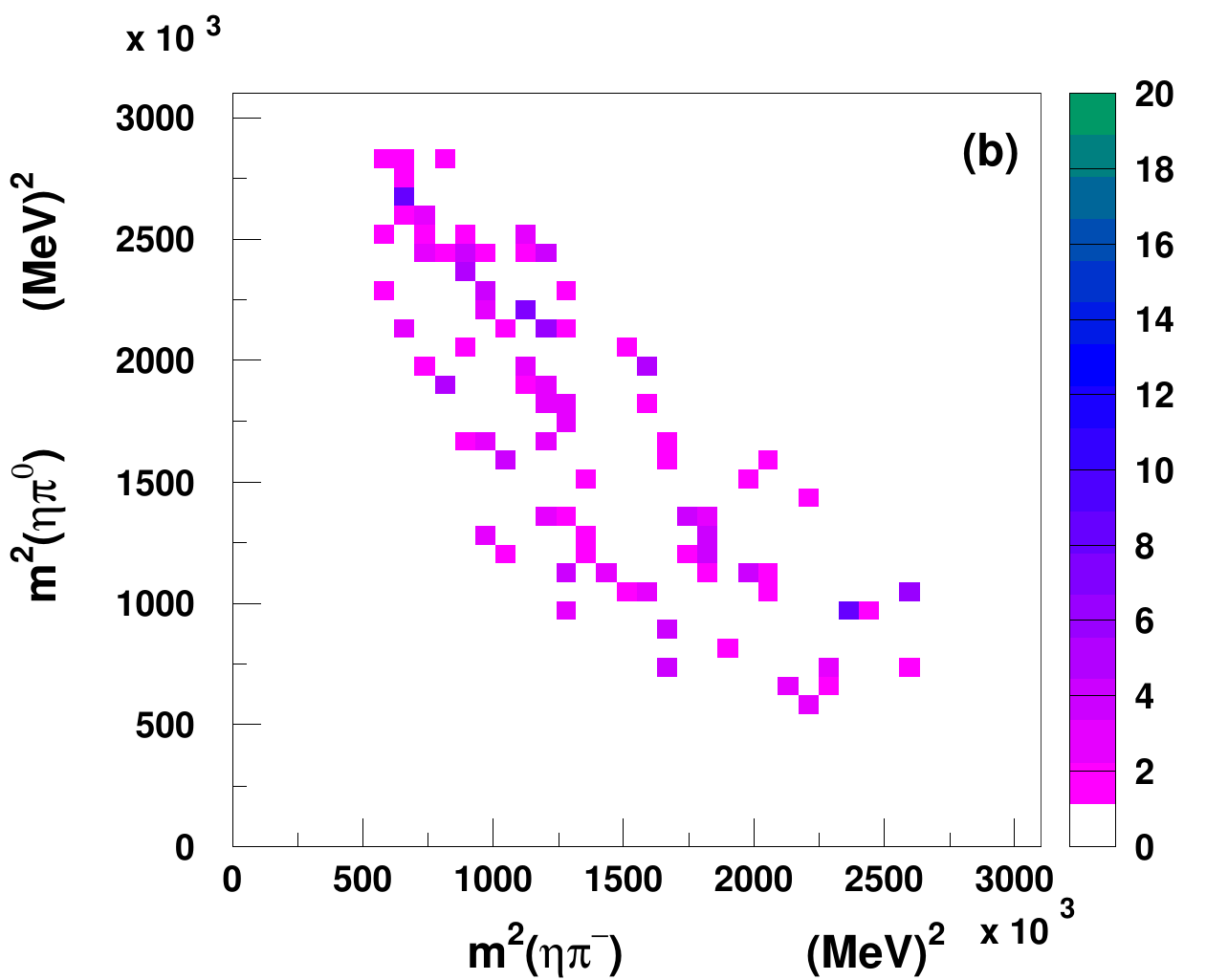} 
\end{tabular}
\caption[]{\label{fig:cbar_chisqr} (Color on line.) 
The difference between the fit and the data in the Dalitz plot 
of $m^{2}(\eta\pi^{0}$ versus $\eta\pi^{-}$ for the reaction $\bar{p}n\rightarrow \eta\pi^{-}\pi^{0}$ 
from the Crystal Barrel experiment~\cite{cbar98}. (a) Does not include the $\pi_{1}(1400)$ while
(b) does include the $\pi_{1}(1400)$. There are clear systematic discrepancies present in (a) that
are not present when the $\pi_{1}(1400)$ is included.}
\end{figure}

The exotic $\pi_{1}$ state was confirmed by the Crystal Barrel Experiment which studied
antiproton-neutron annihilation at rest in the reaction 
$\bar{p}n\rightarrow \eta \pi^{-} \pi^{0}$~\cite{cbar98}. The Dalitz plot for this final
state is shown in Figure~\ref{fig:cbar_p1_1400} where bands for the $a_{2}(1320)$ and 
$\rho(770)$ are clearly seen. They reported a $1^{-+}$ state with a mass of 
$1.40\pm 0.020 \pm 0.020$~GeV and a width of $0.310\pm 0.050^{+0.050}_{-0.030}$~GeV.
While the signal is not obvious in the Dalitz plot, if one compares the difference between 
a fit to the data without and with the $\pi_{1}(1400)$, a clear discrepancy is seen when
the $\pi_{1}(1400)$ is not included (see Figure~\ref{fig:cbar_chisqr}). While the $\pi_{1}(1400)$ 
was only a small fraction of the $a_{2}(1320)$ in the E852 measurement~\cite{Thompson:1997bs}, 
Crystal Barrel observed the two states produced with comparable strength. 

Crystal Barrel also studied the reaction $\bar{p}p\rightarrow \eta \pi^{0} \pi^{0}$~\cite{cbar99}. 
Here, a weak signal was observed for the $\pi_{1}(1400)$ (relative to the $a_{2}(1320)$) with 
a mass of $1.360\pm 0.025$~GeV and a width of $0.220\pm 0.090$~GeV. In $I=0$ $\bar{p}p$
annihilations, the $a_{2}(1320)$ is produced strongly from the $^{1}S_{0}$ atomic state. 
However, $\bar{p}n$ is isospin $1$ and $^{1}S_{0}$ state is forbidden. Thus, the strong $a_{2}$ 
production from $\bar{p}p$ is suppressed in $\bar{p}d$ annihilations---making the $\pi_{1}(1400)$
production appear enhanced relative to the $a_{2}(1320)$ in the latter reaction. 

An analysis by the E852-IU group of data for the reaction $\pi^{-}p\rightarrow n \eta \pi^{0}$
found evidence for the exotic $1^{-+}$ partial wave, but were unable to describe it as
a Breit-Wigner-like $\pi_{1}(1400)$  $\eta\pi^{0}$ resonance~\cite{dzierba03}. 
However, a later analysis by the E852 collaboration of the same final state and data confirmed 
their earlier observation of the $\pi_{1}(1400)$~\cite{Adams:2006sa}. E852 found a mass of 
$1.257\pm 0.020\pm 0.025$~GeV and a width of $0.354\pm 0.064\pm 0.058$~GeV with 
the $\pi_{1}(1400)$ produced via natural parity exchange ($M^{\epsilon}=1^{+}$).
Much of the discrepancy between these two works arise from the treatment of backgrounds.
The E852 collaboration consider no background phase, and attribute all phase motion to
resonances. The E852-IU group allow for non-resonant interactions in the exotic channel,
these background processes are sufficient to explain the observed phase motion.

The $\pi_{1}(1400)$ was also reported in $\bar{p}p$ annihilation into four-pion final states
by both Obelix~\cite{Salvini-04} and Crystal Barrel~\cite{Duenweber-04} (conference 
proceedings only).  They both observed the $\pi_{1}(1400)$ decaying to $\rho\pi$ final 
states, however there is some concern about the production mechanism. The $\eta\pi$ signal 
arises from annihilation from p-wave initial state, while the signal in $\rho\pi$ come 
from the $^{1}S_{0}$ initial state. Thus, it is unlikely that the exotic state seen in $\eta\pi$ and 
that seen in $\rho\pi$ are the same. The origin of these may not be due to an exotic 
resonance, but rather some re-scattering mechanism that has not been properly accounted for.

Interpretation of the $\pi_{1}(1400)$ has been problematic. Its mass is lower than most
predicted values from models, and its observation in only a single decay mode ($\eta\pi$) is not
consistent with models of hybrid decays. Donnachie and Page showed that the $\pi_{1}(1400)$
could be an artifact of the production dynamics. They demonstrated that is possible to 
understand the $\pi_{1}(1400)$ peak as a consequence of the $\pi_{1}(1600)$ (see 
Section~\ref{sec:pi1_1600}) interfering with a non-resonant Deck-type background with 
an appropriate relative phase~\cite{Donnachie:1998ya}.
Zhang~\cite{Zhang:2001sb} considered a molecular picture where the $\pi_{1}(1400)$ was 
an $\eta(1295)\pi$ molecule. However, the predicted decays were inconsistent with
the observations of the $\pi_{1}(1400)$.

Szczepaniak~\cite{Szczepaniak:2003vg} considered a model in which $t$-channel forces
could give rise to a background amplitude which could be responsible for the observed
$\pi_{1}(1400)$. In his model, meson-meson interactions which respected chiral symmetry
were used to construct the $\eta\pi$ $p$-wave interaction much like the $\pi\pi$ $s$-wave
interaction gives rise to the $\sigma$ meson. They claimed that the $\pi_{1}(1400)$ 
was not a QCD bound state, but rather dynamically generated by meson exchange forces. 

Close and Lipkin noted that because the SU(3) multiplets to which a hybrid 
and a multiquark state belong are different, that the $\eta\pi$ and $\eta^{\prime}\pi$ decays
might be a good way to distinguish them. They found that for a multiquark state, the 
$\eta\pi$ decay should be larger than $\eta^{\prime}\pi$, while the reverse is true for a
hybrid meson~\cite{Close:1987aw}. A similar observation was made by Chung~\cite{Chung:2002fz}
who noted that in the limit of the $\eta$ being a pure octet state,
that the decay of an octet $1^{-+}$ state to an $\eta\pi$ $p$-wave is forbidden. Such a 
decay can only come from a decuplet state. Given that the pseudoscalar mixing angle for
the  $\eta$ and $\eta^{\prime}$ are close to this assumption, they argue that the $\pi_{1}(1400)$
is $qq\bar{q}\bar{q}$ in nature. 

While the interpretation of the $\pi_{1}(1400)$ is not clear, most analyses agree that
there is intensity in the $1^{-+}$ wave near this mass. A summary of all reported 
masses and widths for the $\pi_{1}(1400)$ are given in Table~\ref{tab:pi1400}. All 
are reasonably consistent, and even the null observations of VES and E852-IU all concur 
that there is strength near $1.4$~GeV in the $J^{PC}$ exotic wave.  However The E852 and 
VES results can be explained as either non-resonant background~\cite{Szczepaniak:2003vg}, 
or non-resonant deck amplitudes~\cite{Donnachie:1998ya}. An other possibility is the opening 
of meson-meson thresholds, such as $f_{1}(1285)\pi$. Unfortunately, no comparisons of these 
hypothesis have been made with the $\bar{p}N$ data (owing to the lack of general availability of
 the data sets), so it is not possible to conclude that they would also explain those data. However,
in our minds, we believe that the evidence favors a non-resonant interpretation of the exotic
$1^{-+}$ signal and that the $\pi_{1}(1400)$ does not exist.

\begin{table}[h!]\centering
\begin{tabular}{crrcc} \hline\hline
Mode & Mass (GeV) & Width (GeV) & Experiment & Reference \\ 
\hline
$\eta\pi^{-}$  & $1.405\pm 0.020$ & $0.18\pm 0.02$ & GAMS & \cite{alde-88} \\
$\eta\pi^{-}$  & $1.343\pm 0.0046$ & $0.1432\pm 0.0125$ & KEK & \cite{aoyagi-93} \\
$\eta\pi^{-}$ & $1.37\pm 0.016$ & $0.385\pm 0.040$ & E852 &\cite{Thompson:1997bs} \\
$\eta\pi^{0}$ & $1.257\pm0.020$ &$0.354\pm 0.064$ & E852 &\cite{Adams:2006sa} \\
$\eta\pi$      & $1.40\pm 0.020$ & $0.310\pm 0.050$ & CBAR & ~\cite{cbar98} \\
$\eta\pi^{0}$  & $1.36\pm 0.025$ & $0.220\pm 0.090$ & CBAR & ~\cite{cbar99} \\
$\rho\pi$      & $1.384\pm 0.028$ & $0.378\pm 0.058$ & Obelix & \cite{Salvini-04} \\
$\rho\pi$      & $\sim 1.4$ & $\sim 0.4$ & CBAR & \cite{Duenweber-04} \\
$\eta\pi$      & $1.351\pm 0.030$ & $0.313\pm 0.040$ & PDG & \cite{amsler08} \\
\hline\hline
\end{tabular}
\caption[]{\label{tab:pi1400}Reported masses and widths of the $\pi_{1}(1400)$ from the
GAMS, KEK, E852, Crystal Barrel (CBAR) and Obelix experiment. Also reported is the 2008 
PDG average for the state.}
\end{table}

\subsection{\label{sec:pi1_1600}The $\pi_{1}(1600)$}
While the low mass, and single observed decay mode, of the $\pi_{1}(1400)$ have presented some
problems in understanding its nature, a second $J^{PC}=1^{-+}$ state is less problematic.
The $\pi_{1}(1600)$ has been observed in diffractive production using incident $\pi^{-}$
beams where its mass and width have been reasonably stable over several experiments and
the decay modes. It may also have been observed in $\bar{p}p$ annihilation. Positive results 
have been reported from VES, E852, COMPASS and others. These are discussed below in 
approximate chronological order.

In addition to their study of the $\eta\pi^{-}$ system, the VES collaboration 
also examined the $\eta^{\prime}\pi^{-}$ system. Here they observed a $J^{PC}=1^{-+}$ 
partial wave with intensity peaking at a higher mass than the $\pi_{1}(1400)$~\cite{Beladidze:93}.
 However, as with the $\eta\pi^{-}$ system, they did not claim the discovery of an 
exotic-quantum-number resonance. VES later reported a combined study of the $\eta^{\prime}\pi^{-}$, 
$f_{1}\pi^{-}$ and $\rho^{0}\pi^{-}$ final states~\cite{Gouz:1992fu}, and reported a 
``resonance-like structure'' with a mass of $1.62\pm 0.02$~GeV and a width of 
$0.24\pm 0.05$~GeV decaying into $\rho^{0}\pi^{-}$. They also noted that the wave with 
$J^{PC}=1^{-+}$ dominates in the $\eta^{\prime}\pi^{-}$ final state, peaking near $1.6$~GeV and 
observed a small $1^{-+}$ signal in the $f_{1}\pi^{-}$ final state.
\begin{figure}[h!]\centering
\includegraphics[width=0.49\textwidth]{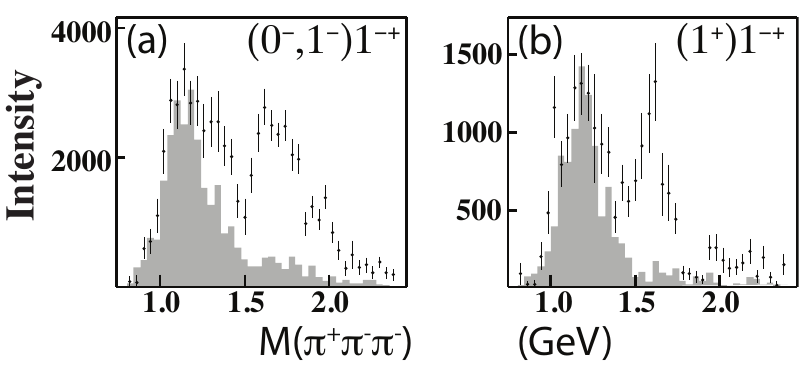}
\caption[]{\label{fig:e852ex}The production of the $1^{-+}$ partial wave as seen in the 
$\pi^{+}\pi^{-}\pi^{-}$ final state by E852. (a) shows the unnatural parity exchange 
($M^{\epsilon}=0^{-}$,$1^{-}$) while (b) shows the natural parity exchange ($M^{\epsilon}=1^{+}$).
(Figure reproduced from reference ~\cite{Adams:1998ff}.)}
\end{figure}

\begin{figure}[h!]\centering
\includegraphics[width=0.47\textwidth]{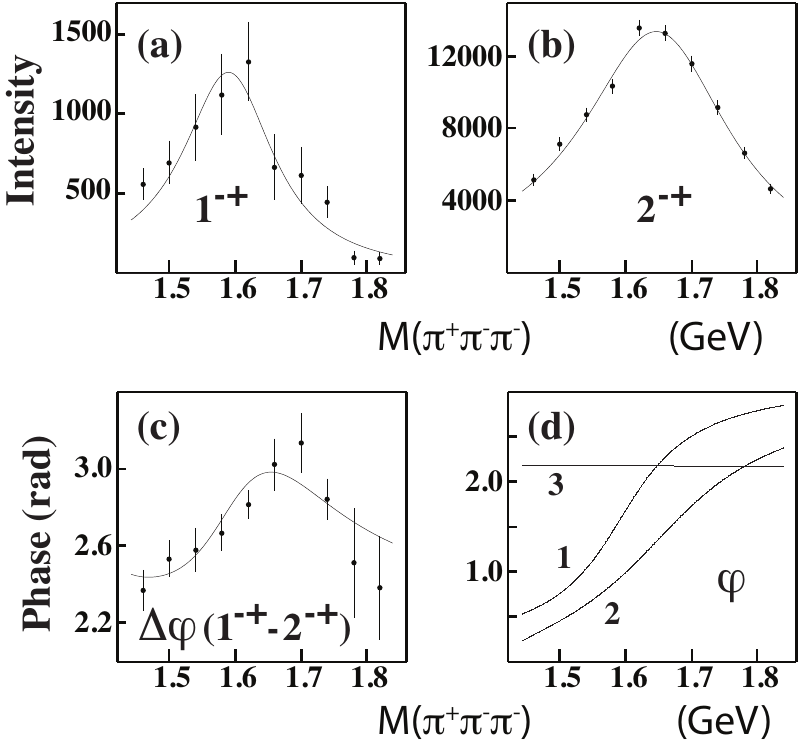}
\caption[]{\label{fig:e8523pi} The results of a PWA to the $\pi^{+}\pi^{-}\pi^{-}$ final
state from E852. (a) shows the intensity of the $J^{PC}=1^{-+}$ wave, (b) shows the 
$2^{-+}$ and (c) shows the phase difference between the two. The solid curves are fits
to two interfering Breit-Wigner distributions. In (d) are shown the phases of the two
Breit-Wigner distributions and (1,2) and a flat background phase (3) that combine to
make the curve in (c).
(Figure reproduced from reference ~\cite{Adams:1998ff}.)}
\end{figure}

Using an $18$~GeV/c $\pi^{-}$ beam incident on a proton target, the E852 collaboration 
carried out a partial wave analysis of the $\pi^{+}\pi^{-}\pi^{-}$ final 
state~\cite{Adams:1998ff,Chung:02}. They saw both the $\rho^{0}\pi^{-}$ and $f_{2}(1270)\pi^{-}$ 
intermediate states and observed a $J^{PC}=1^{-+}$ state which decayed to $\rho\pi$, the 
$\pi_{1}(1600)$. The $\pi_{1}(1600)$ was produced in both natural and unnatural parity 
exchange ( $M^{\epsilon}=1^{+}$ and $M^{\epsilon}=0^{-}$, $1^{-}$) with apparent similar strengths 
in all three exchange mechanisms (see Figure~\ref{fig:e852ex}). In Ref.~\cite{Chung:02}, they
noted that there were issues with the unnatural exchange production. The signal in the 
$M^{\epsilon}=1^{-}$ wave exhibited very strong model dependence and nearly vanished when
larger numbers of partial waves were included. The signal in the $M^{\epsilon}=0^{-}$
partial wave was stable, but its peak was above $1.7$~GeV. They noted that the 
unnatural-parity exchange is expected to die off at higher energies, so their results are 
not at odds with those of VES (see below), where natural parity exchange dominates. 
Even in their data, the unnatural parity exchange waves make
up a small fraction of the total signal. In unnatural parity exchange, they found no significant 
waves, which made a study of phase motion of the $1^{-+}$ in this sector problematic.  
Thus, in their analysis, they only considered the natural parity exchange. There, they found the 
$\pi_{1}(1600)$ to have a mass of $1.593\pm 0.08^{+0.029}_{-0.047}$~GeV and a width of 
$0.168\pm 0.020^{+0.150}_{-0.012}$~GeV. In Figure~\ref{fig:e8523pi} are shown the intensity
of the $1^{-+}$ and $2^{-+}$ ($\pi_{2}(1670)$) partial waves as well as their phase difference.
The phase difference can be reproduced by two interfering Breit-Wigner distributions and
a flat background.

VES  also reported on the $\omega\pi^{-}\pi^{0}$ final state~\cite{Khokholov:00,Zaitsev:00}. In 
a combined analysis of the $\eta^{\prime}\pi^{-}$, $b_{1}\pi$ and $\rho^{0}\pi^{-}$ final states, they reported
the $\pi_{1}(1600)$  state with a mass of $1.61\pm 0.02$~GeV and a width of $0.29\pm 0.03$~GeV
that was consistent with all three final states. To the extent that they observed these states, they
also observed all three final state produced in natural parity exchange ($M^{\epsilon}=1^{+}$). 
They were also able to report relative branching ratios for the three final states as given in 
equation~\ref{eq:pi11600_rates}.
\begin{eqnarray}
\label{eq:pi11600_rates}
b_{1}\pi : 
\eta^{\prime}\pi : 
\rho\pi : & = & 
1: 1\pm 0.3 : 1.5\pm 0.5
\end{eqnarray}
However, there were some issues with the $\rho\pi$ final state. Rather than limiting the rank
of the density matrix as was done in~\cite{Adams:1998ff,Chung:02}, they did not limit it. This allowed
for a more general fit that might be less sensitive to acceptance affects. In this model, they 
did not observe any significant structure in the $1^{-+}$ $\rho\pi$ partial wave above $1.4$~GeV. However,
by looking at how other resonances were produced, they were able to isolate a coherent part
of the density matrix from which they found a statistically significant $1^{-+}$ partial wave
peaking near $1.6$~GeV. While VES was extremely careful not to claim the existence of
the $\rho\pi$ decay of the the $\pi_{1}(1600)$, in the case that it exists, they were able to 
obtain the rates given in equation~\ref{eq:pi11600_rates}.
\begin{figure}[h!]\centering
\includegraphics[width=0.5\textwidth]{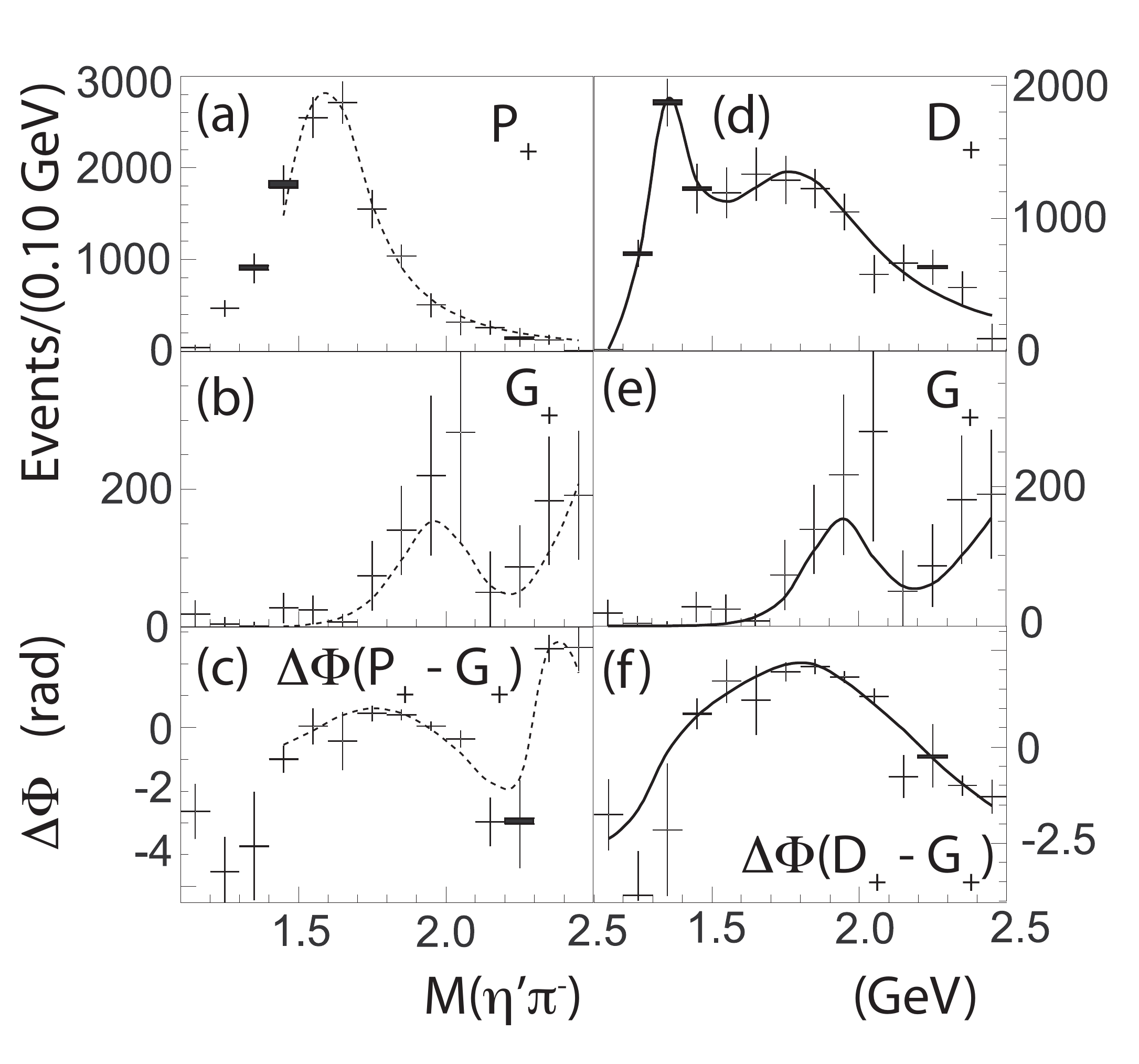}
\caption[]{\label{fig:e852_etaprime} Results from E852 on the $\eta^{\prime}\pi^{-}$ final 
state. (a) shows the $1^{-+}$ partial wave, (b) shows the $4^{++}$ partial wave (an $a_{4}$) 
and (c) shows the phase difference between these. (d) shows the  $2^{++}$ partial wave 
($a_{2}(1320)$), while (e) shows the $a_{4}$ and (f) is the phase difference.
(Figure reproduced from reference~\cite{Ivanov:2001rv}.)}
\end{figure}

In a follow-up analysis, E852 also studied the reaction $\pi^{-}p\rightarrow p\eta^{\prime}\pi^{-}$ to 
examine the  $\eta^{\prime}\pi^{-}$ final state~\cite{Ivanov:2001rv}. They observed, consistent with 
VES~\cite{Beladidze:93}, that the dominant signal was the $1^{-+}$ exotic wave produced 
dominantly in the $M^{\epsilon}=1^{+}$ channel, implying only natural parity exchange. 
They found the signal to be consistent with a resonance, the $\pi_{1}(1600)$ and found
a mass of $1.597\pm 0.010^{+0.045}_{-0.010}$~GeV and a width of $0.340\pm 0.040\pm  
0.050$~GeV. The results of the E852 PWA are shown in Figure~\ref{fig:e852_etaprime}
where the $P_{+}$ wave is the $1^{-+}$, the $D_{+}$ corresponds to the $2^{++}$ $a_{2}$ and
the $G_{+}$ corresponds to the $4^{++}$ $a_{4}$. Clear phase motion is observed between both 
the $2^{++}$ and $4^{++}$ wave and the $1^{-+}$ and the $4^{++}$ wave.

An analysis of Crystal Barrel data at rest for the reaction $\bar{p}p\rightarrow\omega\pi^{+}\pi^{-}\pi^{0}$
was carried by some members of the collaboration~\cite{Baker:2003jh}. They reported evidence
for the $\pi_{1}(1600)$ decaying to $b_{1}\pi$ from both the $^{1}S_{0}$ and $^{3}S_{1}$ initial
states, with the signal being stronger from the former. The total signal including
both initial states, as well as decays with $0$ and $2$ units of angular momentum accounted
for less than $10$\% of the total reaction channel. The mass and width were found consistent
(within large errors) of the PDG value, and only results with the mass and width fixed to the
PDG values were reported. Accounting for the large rate of annihilation to $\omega\pi^{+}\pi^{-}\pi^{0}$
of $13$\%, this would imply that $\bar{p}p\rightarrow \pi_{1}(1600)\pi$ accounts for several
percent of all $\bar{p}p$ annihilations. 

E852 also looked for the decays of the $\pi_{1}(1600)$ to $b_{1}\pi$ and $f_{1}\pi$. The latter
was studied in the reaction $\pi^{-}p\rightarrow p \eta \pi^{+}\pi^{-}\pi^{-}$ with the $f_{1}$
being reconstructed in its $\eta\pi^{+}\pi^{-}$ decay mode~\cite{Kuhn:2004en}.
The $\pi_{1}(1600)$ was seen via interference with both the $1^{++}$ and $2^{-+}$ partial
waves. It was produced via natural parity exchange ($M^{\epsilon}=1^{+}$) and found
to have a mass of $1.709\pm 0.024\pm 0.041$~GeV and a width of 
$0.403\pm 0.080\pm 0.115$~GeV. A second $\pi_{1}$ state was also observed in this
reaction (see Section~\ref{sec:pi1_2015}). 

The $b_{1}\pi$ final state was studied by looking
at the reaction $\pi^{-}p\rightarrow \omega \pi^{-}\pi^{0}p$, with the $b_{1}$ reconstructed in
its  $\omega\pi$  decay mode~\cite{Lu:2004yn}. The $\pi_{1}(1600)$ was seen interfering
with the $2^{++}$ and $4^{++}$ partial waves. In $b_{1}\pi$, they measured a mass of 
$1.664\pm 0.008\pm 0.010$~GeV and a width of $0.185\pm 0.025\pm 0.028$~GeV
for the $\pi_{1}(1600)$. However, the production mechanism was seen to be a mixture
of both natural and unnatural parity exchange, with the unnatural being stronger. As
with the $f_{1}\pi$, they also observed a second $\pi_{1}$ state decaying to $b_{1}\pi$
(see Section~\ref{sec:pi1_2015}).
\begin{table}[h!]\centering
\begin{tabular}{crc} \hline\hline
final state & production ($M^{\epsilon}$) & dominant \\ \hline
$\rho\pi$               & $0^{-}$, $1^{-}$, $1^{+}$ & npe $\sim$ upe \\
$\eta^{\prime}\pi$    & $1^{+}$ & npe \\
$f_{1}\pi$                & $1^{+}$ & npe \\
$b_{1}\pi$               & $0^{-}$, $1^{-}$, $1^{+}$ & upe $>$ npe \\
\hline\hline
\end{tabular}
\caption[]{\label{tab:e852_prod}The production mechanisms for the $\pi_{1}(1600)$ as seen 
in the E852 experiment. Also shown is whether the natural parity exchange (npe) or the 
unnatural parity exchange (upe) is stronger.}
\end{table}

The fact that E852 observed the $\pi_{1}(1600)$ produced in different production mechanisms,
depending on the final state, is somewhat confusing. A summary of the observed mechanisms 
is given in Table~\ref{tab:e852_prod}. In order to understand the variations in production, 
there either needs to be two nearly-degenerate $\pi_{1}(1600)$s, or there is some 
unaccounted-for systematic problem in some of the analyses.
\begin{figure}[h!]\centering
\includegraphics[width=0.5\textwidth]{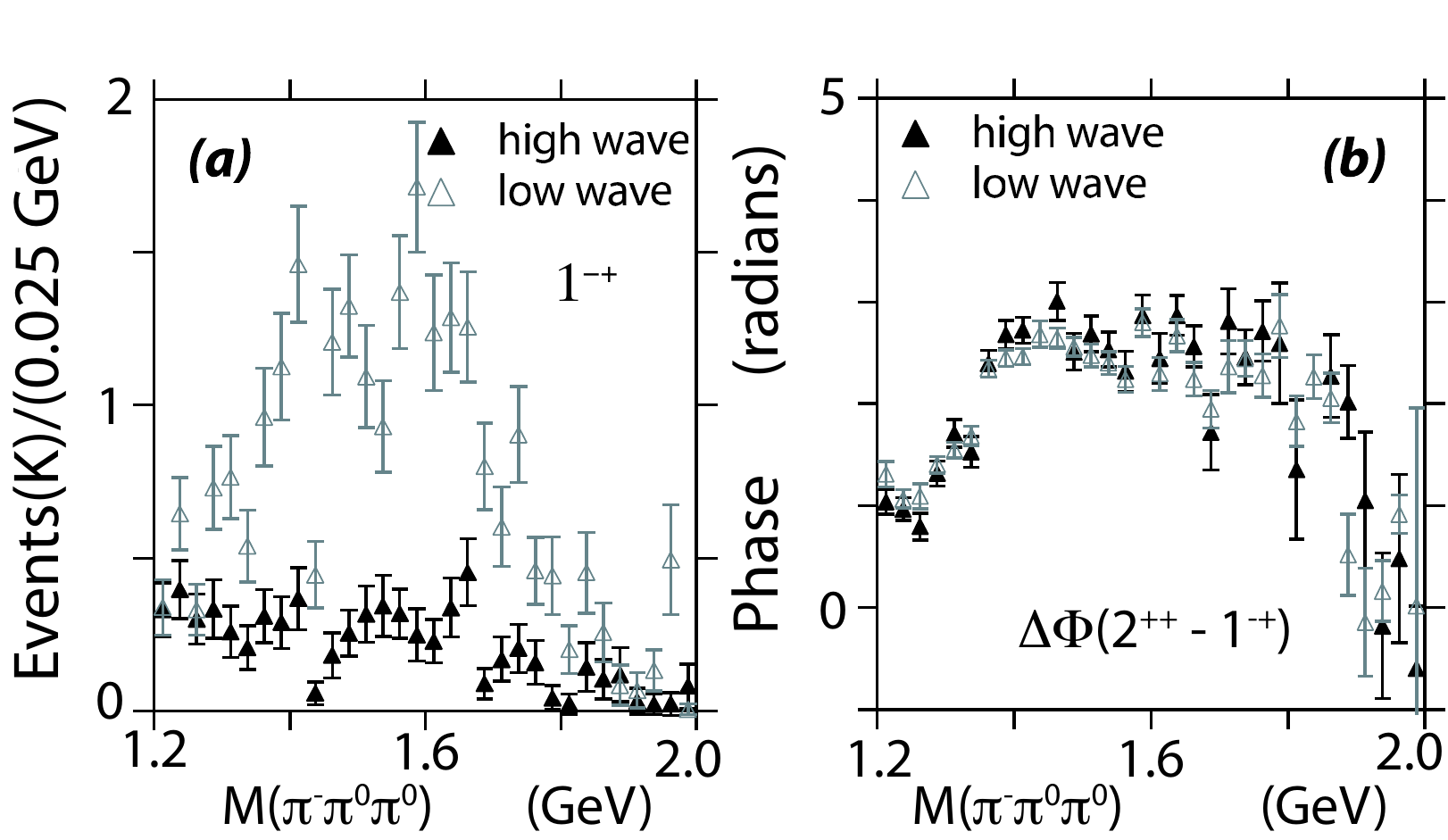}
\caption[]{\label{fig:e852_dz_m00}(Color on line) The PWA solutions for the $1^{-+}$ partial wave for the
$\pi^{-}\pi^{0}\pi^{0}$ final state (a) and its interference with the $2^{++}$ partial wave (b). See text 
for an explanation of the labels. (Figure reproduced from reference~\cite{Dzierba:2005jg}.)}
\end{figure}
\begin{figure}[h!]\centering
\includegraphics[width=0.5\textwidth]{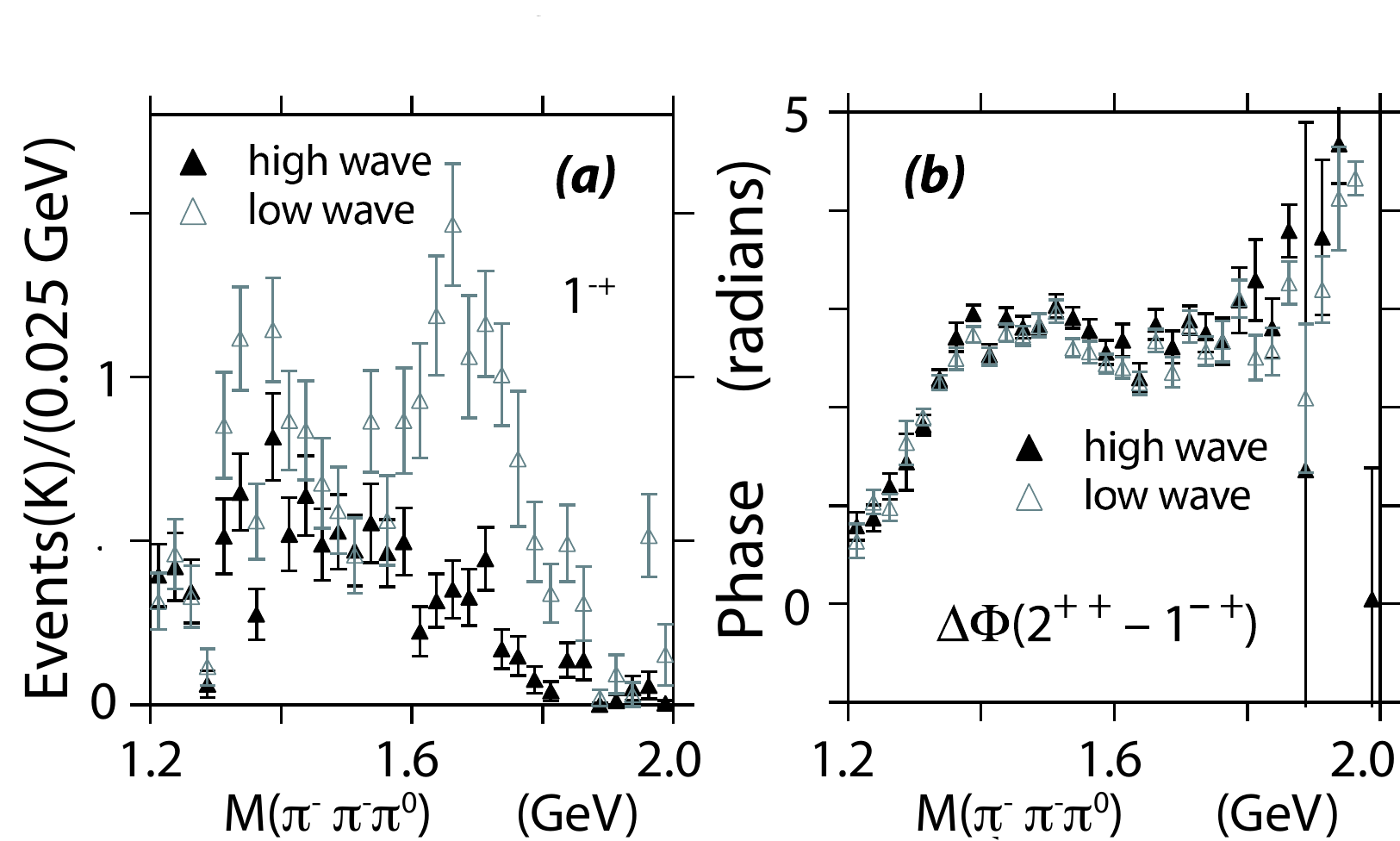}
\caption[]{\label{fig:e852_dz_pmm}(Color on line) The PWA solutions for the $1^{-+}$ partial wave for the
$\pi^{+}\pi^{-}\pi^{-}$ final state (a) and its interference with the $2^{++}$ partial wave (b). See text 
for an explanation of the labels. (Figure reproduced from reference~\cite{Dzierba:2005jg}.)}
\end{figure}

The E852-IU group analyzed an E852 data set that was an order of magnitude larger than that 
used by E852 in Refs.~\cite{Adams:1998ff,Chung:02}. In this larger data set, they looked at the
reactions $\pi^{-}p\rightarrow n \pi^{+}\pi^{-}\pi^{-}$ and $\pi^{-}p\rightarrow n \pi^{-}\pi^{0}\pi^{0}$
and carried out a partial wave analysis for both the $\pi^{+}\pi^{-}\pi^{-}$ and the $\pi^{-}\pi^{0}\pi^{0}$ 
final states. This yielded solutions that were consistent with both final states~\cite{Dzierba:2005jg}.
In this analysis, they carried out a systematic study of which partial waves were important
in the fit. When they used the same wave set as in the E852 analysis~\cite{Adams:1998ff,Chung:02}, 
they found the same solution showing a signal for the $\pi_{1}(1600)$ in both final states. However,
when they allowed for more partial waves, they found that the signal for the $\pi_{1}(1600)$ went
away. Figure~\ref{fig:e852_dz_m00} shows these results for the $\pi^{-}\pi^{0}\pi^{0}$ final state,
while Figure~\ref{fig:e852_dz_pmm} shows the results for the $\pi^{+}\pi^{-}\pi^{-}$ final
state. In both figures, the ``low wave'' solution matches that from E852, while their ``high wave'' 
solution shows no intensity for the $\pi_{1}(1600)$ in either channel. An important point is that
in both their high-wave and low-wave analyses, the phase difference between the exotic $1^{-+}$
wave and the $2^{++}$ wave are the same (and thus the same as in the E852 analysis). While not
shown here, the same is also true for the $1^{-+}$ and $2^{-+}$ waves. 
\begin{table}[h!]\centering
\begin{tabular}{ccccccc}\hline\hline
$\pi_{2}(1670)$ & \multicolumn{2}{c}{$M^{\epsilon}=0^{+}$}
                        & \multicolumn{2}{c}{$M^{\epsilon}=1^{+}$}
                        & \multicolumn{2}{c}{$M^{\epsilon}=1^{-}$} \\
Decay               & L & H & L & H & L & H \\ \hline
$(f_{2}\pi)_{S}$      & $\times$ & $\times$ & $\times$ & $\times$ & $\times$ &     \\
$(f_{2}\pi)_{D}$      & $\times$ & $\times$ & $\times$ & $\times$ &  &  \\
$[(\pi\pi)_{S}]_{D}$ &$\times$ & $\times$ &                 & $\times$ &  &  \\
$(\rho\pi)_{P}$      &$\times$ & $\times$ &                 & $\times$ &  &  \\
$(\rho\pi)_{F}$      &                & $\times$ &                 & $\times$ &  &  \\
$(f_{0}\pi)_{D}$      &                 & $\times$ &                 & $\times$ &  &  \\
\hline\hline
\end{tabular}
\caption[]{\label{tab:p2-decays}The included decays of the $\pi_{2}(1670)$ in two analyses
of the $3\pi$ final state. ``L'' is the wave set used in the E852 analysis~\cite{Adams:1998ff,Chung:02}.
``H'' is the wave set used in the higher statistics analysis~\cite{Dzierba:2005jg}.}
\end{table}

\begin{figure}[h!]\centering
\includegraphics[width=0.35\textwidth]{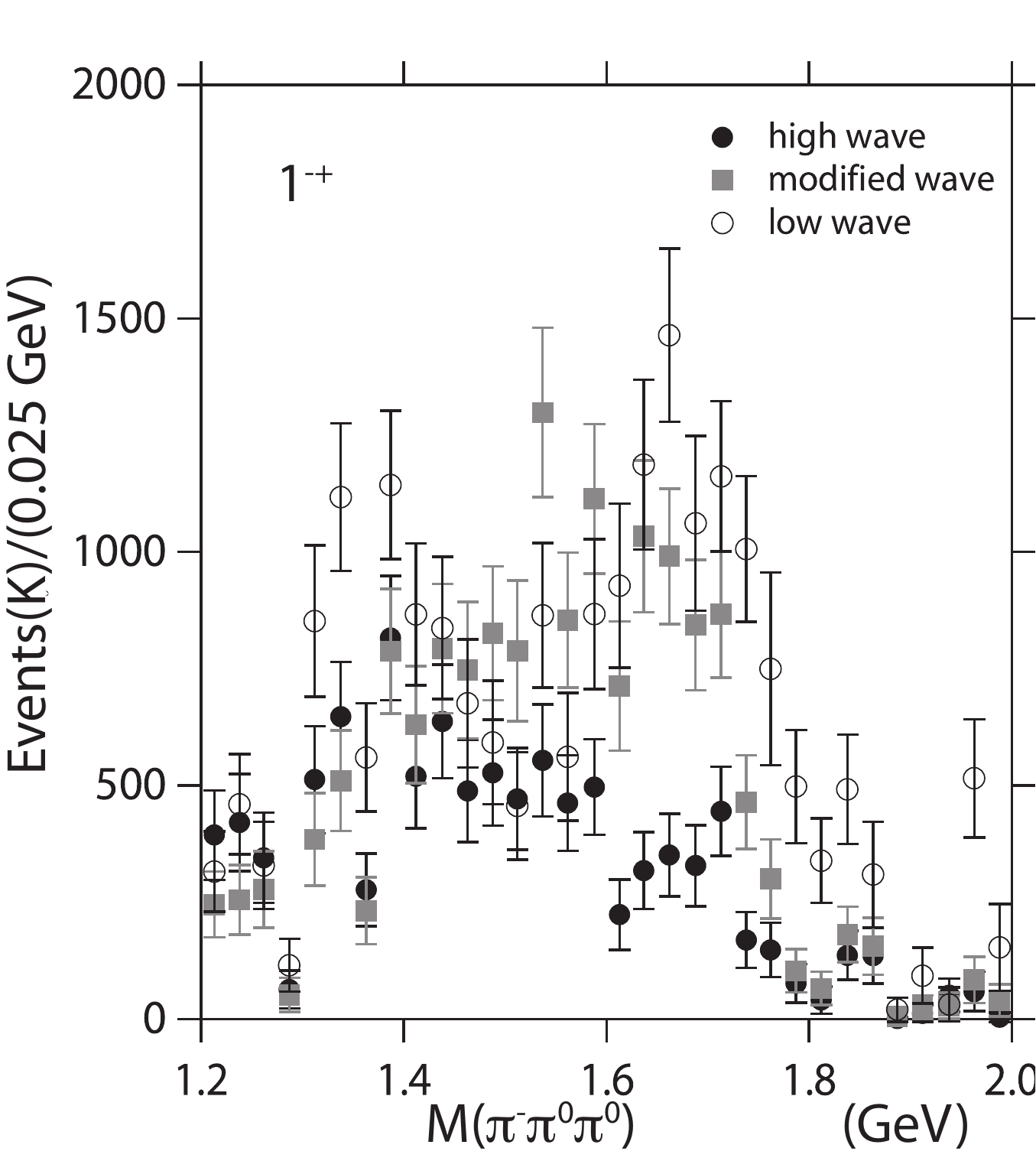}
\caption[]{\label{fig:e852_dz_pi2} The $1^{-+}$ intensity for the charged mode for the high-wave 
set (filled circles), the modified high-wave set (filled squares), and the low-wave set (open circles). 
In the modified high-wave set the two $\rho\pi$ decays of the $\pi_{2}(1670)$ were removed
from the fit. (Figure reproduced from reference~\cite{Dzierba:2005jg}.)}
\end{figure}

E852-IU carried out a study to determine which of the additional waves in their ``high wave'' 
set were absorbing the intensity of the $\pi_{1}(1600)$. They found that the majority of this was 
due to the inclusion of the $\rho\pi$ decay of the $\pi_{2}(1670)$. The partial waves associated with the 
$\pi_{2}(1670)$ in both analyses are listed in Table~\ref{tab:p2-decays}. While the production
from $M^{\epsilon}=0^{+}$ is similar for both analyses, the E852 analysis only included 
the $\pi_{2}(1670)$ decaying to $f_{2}\pi$ in the $M^{\epsilon}=1^{+}$ production.
The high-statistics analysis included both $f_{2}\pi$ and $\rho\pi$ in both production mechanisms.
The PDG~\cite{amsler08} lists the two main decays of the $\pi_{2}(1670)$ as $f_{2}\pi$ (56\%) and
$\rho\pi$ (31\%), so it seems odd to not include this latter decay in an analysis
including the $\pi_{2}(1670)$. Figure~\ref{fig:e852_dz_pi2} shows the results of removing the 
$\rho\pi$ decay from the ``high wave'' set for the $\pi^{+}\pi^{-}\pi^{-}$ final state. This
decay absorbs a significant portion of the $\pi_{1}(1600)$ partial wave.

In the E852-IU analyses, the fact that the phase motion of the $1^{-+}$ exotic wave relative 
to other partial waves agrees with with those differences as measured by E852, and are the same 
in both the high-wave and low-wave analyses is intriguing. This could be interpreted as a 
$\pi_{1}(1600)$ state which is simply absorbed by the stronger $\pi_{2}(1670)$ as more partial 
waves are added. However, given the small actual phase difference between the $1^{-+}$ and 
$2^{-+}$ partial waves (see Figure~\ref{fig:e8523pi}), the opposite conclusion is also possible, 
particularly if some small non-zero background phase were present. Here, the $1^{-+}$ signal 
is due to leakage from the stronger $\pi_{2}$ and no $\pi_{1}(1600)$ is needed in the $\rho\pi$ 
final state. 

\begin{figure}[h!]\centering
\includegraphics[width=0.5\textwidth]{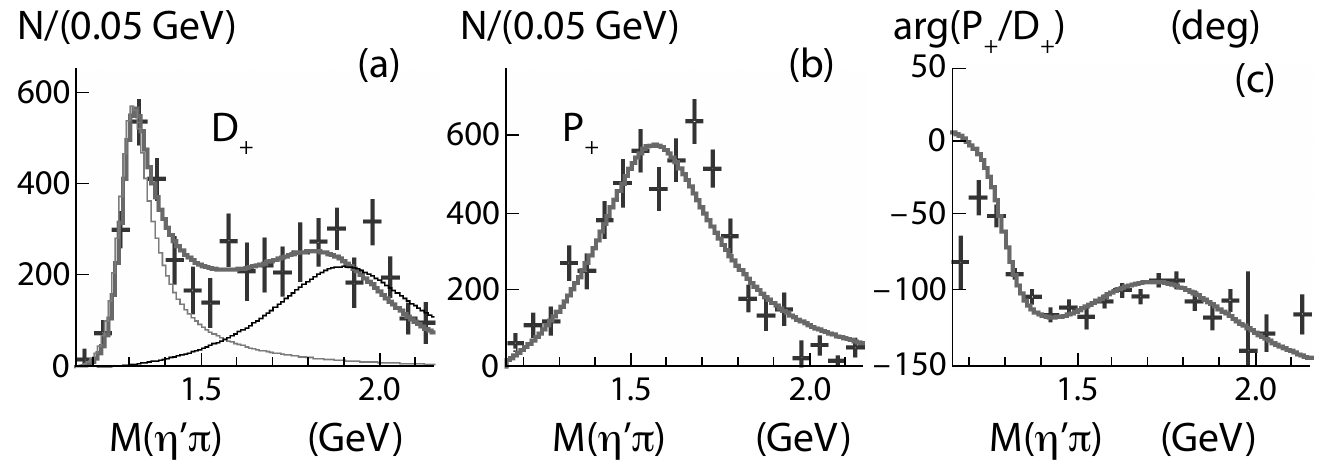}
\caption[]{\label{fig:ves_etaprpi} The results of a partial wave analysis on the $\eta^{\prime}\pi^{-}$
final state from VES. (a) shows he $2^{++}$ partial wave in $\omega\rho$, (b) shows the 
$1^{-+}$ partial wave in $b_{1}\pi$ and (c) shows the interference between them.
(Figure reproduced from reference~\cite{Amelin:2005ry}.)}
\end{figure}

\begin{figure}[h!]\centering
\includegraphics[width=0.5\textwidth]{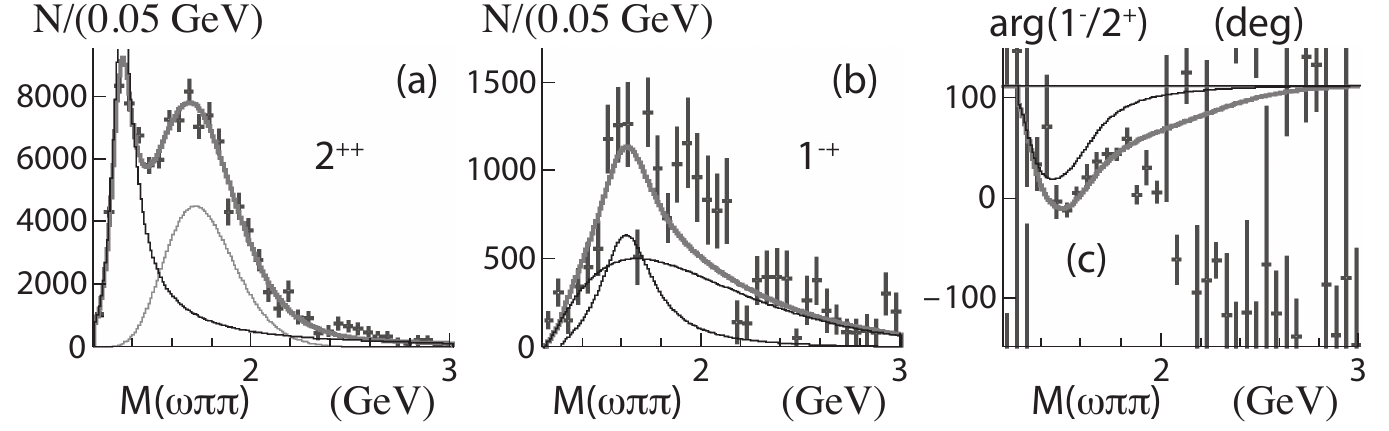}
\caption[]{\label{fig:ves_b1pi}The results of a partial wave analysis on the $b_{1}\pi$
final state from VES. (a) shows he $2^{++}$ partial wave, (b) shows the $1^{-+}$ partial wave and
(c) shows the interference between them.
(Figure reproduced from reference~\cite{Amelin:2005ry}.)}
\end{figure}

\begin{figure}[h!]\centering
\includegraphics[width=0.5\textwidth]{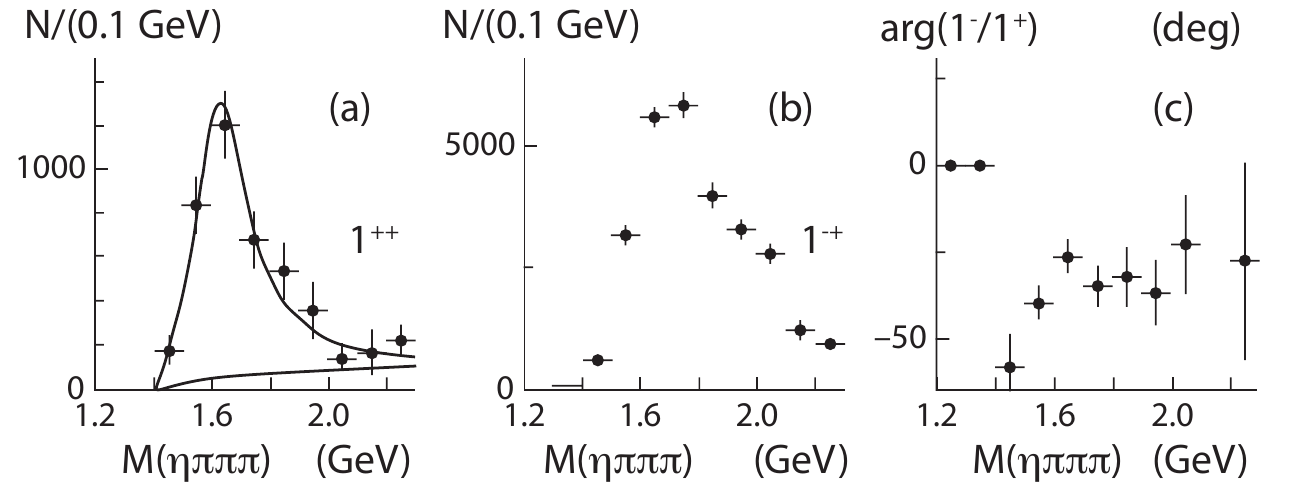}
\caption[]{\label{fig:ves_f1pi}The results of a partial wave analysis on the $f_{1}\pi$
final state from VES. (a) shows he $1^{++}$ partial wave, (b) shows the $1^{-+}$ partial wave and
(c) shows the interference between them.
(Figure reproduced from reference~\cite{Amelin:2005ry}.)}
\end{figure}

The VES results have been summarized in a review of all their work on hybrid 
mesons~\cite{Amelin:2005ry}. This included an updated summary of the $\pi_{1}(1600)$
in all four final states, $\eta^{\prime}\pi$ $\rho\pi$, $b_{1}\pi$ and  $f_{1}\pi$. 
In the $\eta^{\prime}\pi$ final state (Figure~\ref{fig:ves_etaprpi}), 
they note that the $1^{-+}$
wave is dominant. While they were concerned about the nature of the higher-mass part of
the $2^{++}$ spectrum ($a_{2}(1700)$ or background) they find that a resonant description of 
$\pi_{1}(1600)$ is possible in both cases. For the case of the $b_{1}\pi$ final state
(Figure~\ref{fig:ves_b1pi}), they find that
the contribution of a $\pi_{1}(1600)$ resonance is required. In a combined fit to both the 
$\eta^{\prime}\pi$ and $b_{1}\pi$ data, they find a mass of $1.56\pm 0.06$~GeV and a width 
of $0.34\pm 0.06$~GeV for the $\pi_{1}(1600)$. In the $f_{1}\pi$ final state
(Figure~\ref{fig:ves_f1pi}),  they find a
resonant description of the the $\pi_{1}(1600)$ with a mass of $1.64\pm 0.03$~GeV and
a width of $0.24\pm 0.06$~GeV which they note is compatible with their measurement
in the previous two final states. They also note, that in contradiction with E852~\cite{Kuhn:2004en},
they find no significant $1^{-+}$ intensity above a mass of $1.9$~GeV (see Section~\ref{sec:pi1_2015}).
For the $\rho\pi$ final state, they are unable to conclude that the $\pi_{1}(1600)$ is present.

They note that the partial-wave analysis of the $\pi^{+}\pi^{-}\pi^{-}$ system finds a significant
contribution from the $J^{PC}=1^{−+}$ wave in the $\rho\pi$ channel ($2$ to $3$\% of the total intensity).
Some of the models in the partial-wave analysis of the exotic wave lead to the appearance of a peak
near a mass of $1.6$~GeV which resembles the $\pi_{1}(1600)$. However, the dependence of the 
size of this peak on the model used is significant~\cite{Zaitsev:00}. They note that because the 
significance of the wave depends very strongly on the assumptions of coherence used in the analysis,
the results for $3\pi$ final states on the nature of the $\pi_{1}(1600)$ are not reliable.

To obtain a limit on the branching fraction of $\pi_{1}(1600)$ decay to $\rho\pi$, they  looked at 
their results of the production of the $\pi_{1}(1600)$ in the charge-exchange reaction to 
$\eta^{\prime}\pi^{0}$ versus that of the $\eta^{\prime}\pi^{-}$ final state. They conclude that the
presence of the $\pi_{1}(1600)$ in $\eta^{\prime}\pi^{-}$ and its absence in $\eta^{\prime}\pi^{0}$
preclude the formation of the $\pi_{1}(1600)$ by $\rho$ exchange. From this, they obtain the 
relative branching ratios for the $\pi_{1}(1600)$ as given in equation~\ref{eq:pi11600_rates3}.
\begin{eqnarray}
\label{eq:pi11600_rates3}
b_{1}\pi : 
f_{1}\pi :
\rho\pi : 
\eta^{\prime}\pi & = & 
1.0 \pm .3: 1.1 \pm .3 : < .3 :1
\end{eqnarray}

While the results on $\rho\pi$ between E852 and VES seem at odds, we believe that these 
discrepancies are the result of the assumptions made in the analyses. These assumptions
then manifest themselves in the interpretation of the results. The VES analyses fit both the 
real and imaginary parts of their amplitudes independently. However, for analytic functions, 
the two parts are not independent. Not using these constraints can lead to results that may 
be unphysical, and at a minimum, are discarding important constraints on the amplitudes.
All of which can lead to difficulties in interpreting the results. In E852, many of their results 
rely on the assumption of a flat background phase, but there are many examples where 
this is not true. Thus, their results are biased towards a purely resonant description of the 
data, rather than a combination of resonant and non-resonant parts. It is also somewhat
disappointing that E852 is unable to make statements about relative decay rates, or carry
out a coupled channel analysis of their many data sets. Our understanding is that this 
is due to issues in modelling the rather tight trigger used in collecting their data.

The CLAS experiment at Jefferson Lab studied the reaction 
$\gamma p\rightarrow \pi^{+}\pi^{+}\pi^{-} (n)_{miss}$ to look for the production of the 
$\pi_{1}(1600)$~\cite{Nozar:09}. The photons were produced by bremsstrahlung from a 
$5.7$~GeV electron beam. While there was significant contributions from Baryon resonances
in their data, they attempted to remove this by selective cuts on various kinematic regions.
The results of their partial-wave analysis show clear signals for the $a_{1}(1270)$, the 
$a_{2}(1320)$ and the $\pi_{2}(1670)$, but show no evidence for the $\pi_{1}(1600)$ decaying 
into three pions. They place and upper limit of the production and subsequent decay of
the $\pi_{1}(1600)$ to be less than 2\% of the $a_{2}(1320)$. There results imply that the
$\pi_{1}(1600)$ is not strongly produced in photoproduction, the $\pi_{1}(1600)$ does not
decay to $3\pi$, or both.

\begin{figure}[h!]\centering
\includegraphics[width=0.45\textwidth]{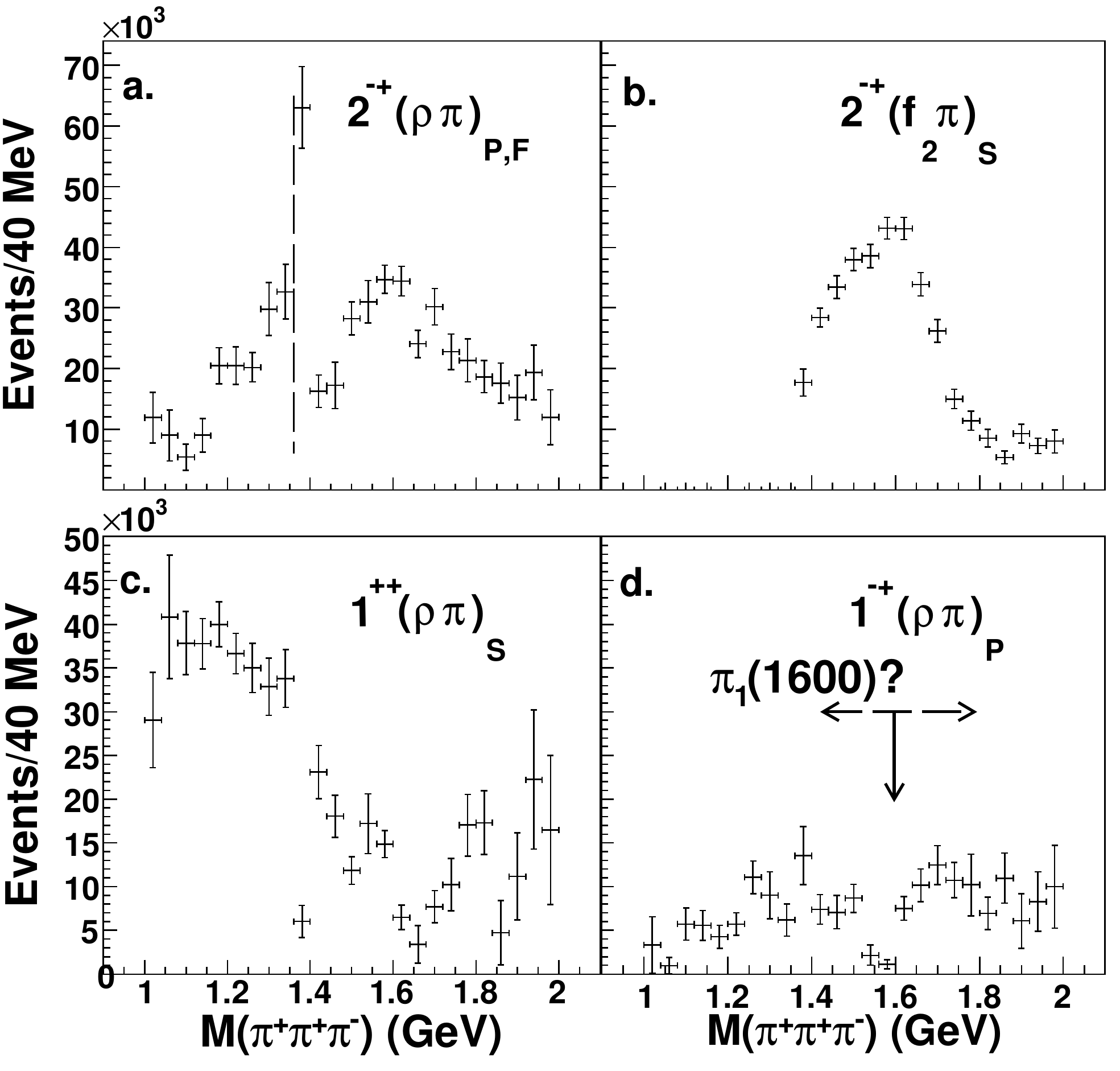}
\caption[]{\label{fig:clas3pi} The results from CLAS of a partial-wave analysis photoproduction
data of the $3\pi$ final state. Intensity is seen in the $2^{-+}$ partial wave, (a) and (b), as
well as the $1^{++}$ partial wave (c). In the $1^{-+}$ exotic wave, (d), no intensity is observed. 
(Figure reproduced from reference~\cite{Nozar:09}.)}
\end{figure}

\begin{figure}[h!]\centering
\includegraphics[width=0.45\textwidth]{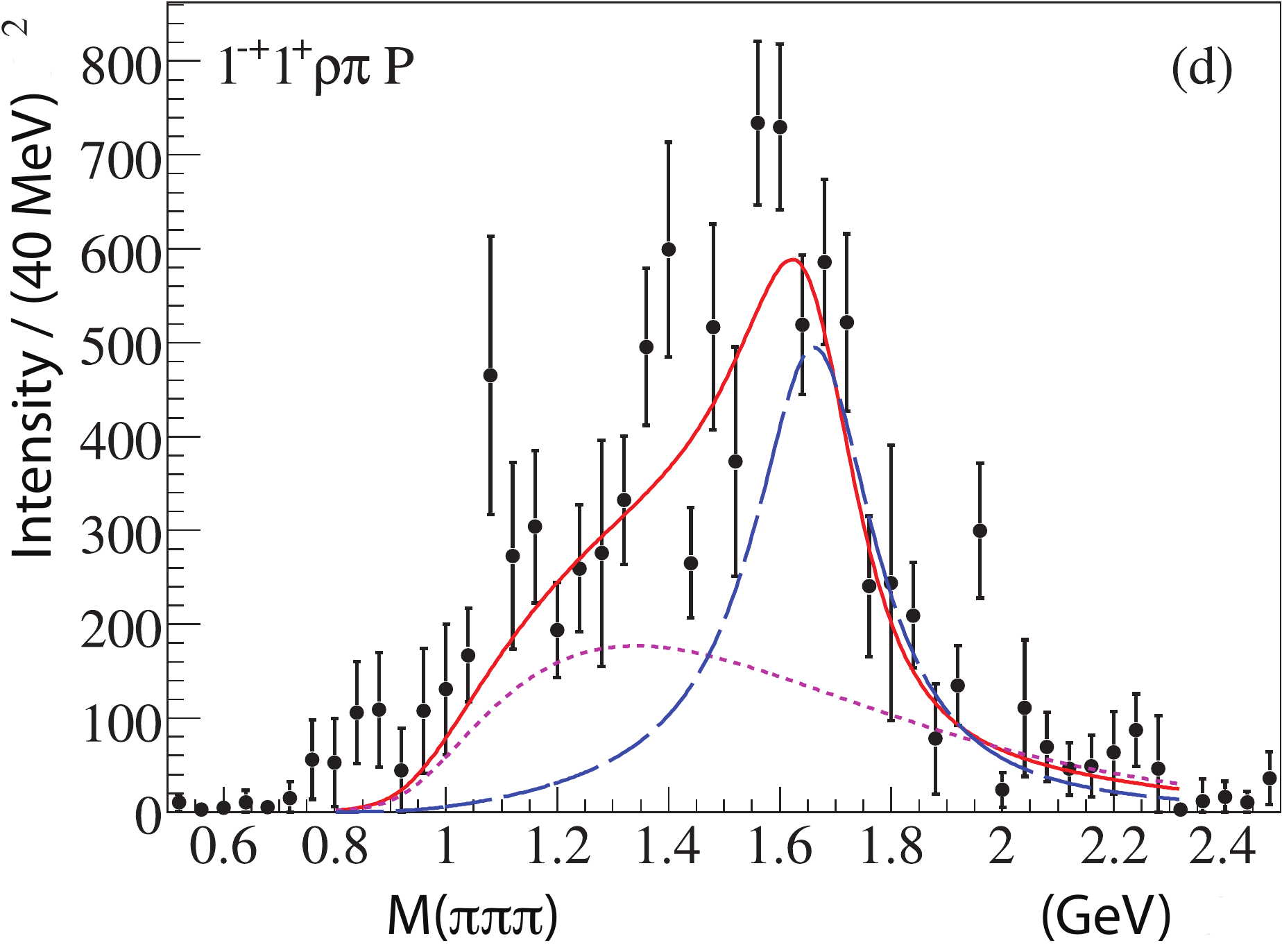} 
\caption[]{\label{fig:compass_int} (Color on line.) COMPASS results showing the intensity of 
the exotic $1^{-+}$ wave. The solid (red) curve shows a fit to the corresponding resonances. 
The dashed (blue) curve is the $\pi_{1}(1600)$ while the dotted (magenta) curve is background.
(Figure reproduced from reference~\cite{Alekseev:2009xt}.)}
\end{figure}

\begin{figure}[h!]\centering
\includegraphics[width=0.50\textwidth]{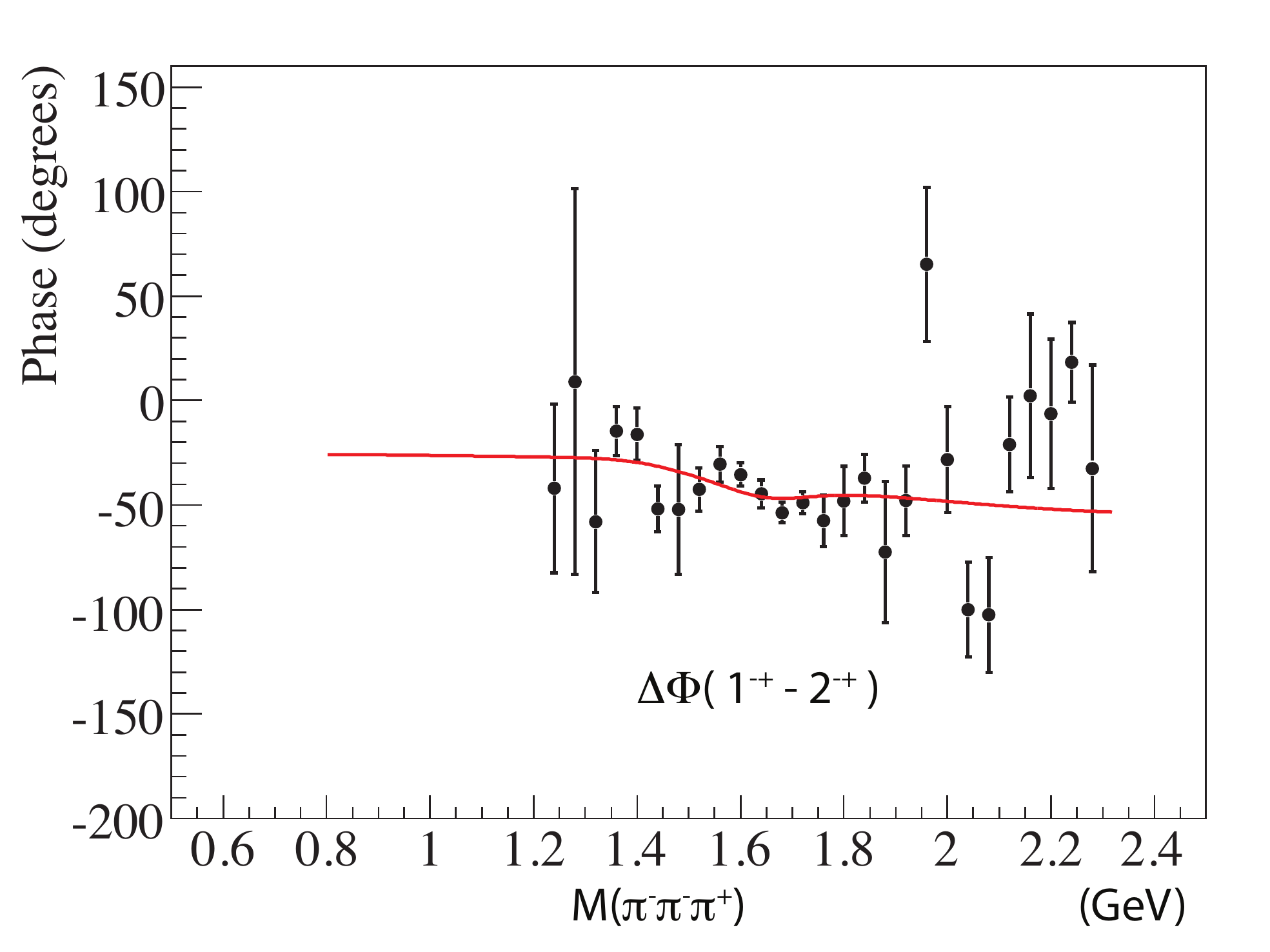}
\caption[]{\label{fig:compass_phs} (Color on line.) COMPASS results showing the phase  
difference between the exotic $1^{-+}$ wave and the $2^{-+}$ wave. The solid (red) curve 
shows a fit to the corresponding resonances.
(Figure reproduced from reference~\cite{Alekseev:2009xt}.)}
\end{figure}

The COMPASS experiment has reported their first study of the diffractively produced 
$3\pi$ final state~\cite{Alekseev:2009xt,Grube:2010}.  They used a $190$~GeV/c beam of pions 
to study the reaction $\pi^{-} Pb \rightarrow \pi^{-}\pi^{-}\pi^{+} X$. In their partial-wave
analysis of the $3\pi$ final state, they observed the $\pi_{1}(1600)$ with a mass of 
$1.660\pm 0.010^{+0}_{-0.064}$~GeV and a width of $0.269\pm 0.021^{+0.042}_{-0.064}$~GeV.
The $\pi_{1}(1600)$ was produced dominantly in natural parity exchange ($M^{\epsilon}=1^{+}$)
although unnatural parity exchange also seemed to be required. However, the level was not
reported. The wave set (in reference~\cite{Grube:2010}) used appears to be somewhat larger 
than that used in the high-statistics study of E852-IU~\cite{Dzierba:2005jg}.  Thus, in
the COMPASS analysis, the $\rho\pi$ decay of the $\pi_{2}(1670)$ does not appear to absorb 
the exotic intensity in their analysis. They also report on varying the rank of the fit with
the $\pi_{1}(1600)$  and the results being robust against these changes. One point of small concern is that 
the mass and width that they extract for the $\pi_{1}(1600)$ are essentially identical to those 
for the $\pi_{2}(1670)$. For the latter, they observed  a mass of $1.658\pm 0.002^{+0.024}_{-0.008}$~GeV 
and a width of $0.271\pm 0.009^{+0.022}_{-0.024}$~GeV. However, the strength of the exotic wave
appears to be about $20$\% of the $\pi_{2}$, thus feed through seems unlikely. Results from their 
partial-wave analysis are shown in Figures~\ref{fig:compass_int} and~\ref{fig:compass_phs}. 
These show the $1^{-+}$ partial wave and the phase difference between the $1^{-+}$ and $2^{-+}$
waves. The solid curves are the results of mass-dependent fits to the $\pi_{1}(1600)$ and 
$\pi_{2}(1670)$. 

Table~\ref{tab:p1_1600} summarizes the masses and widths found for the $\pi_{1}(1600)$
in the four decay modes and from the experiments which have seen a positive result.
While the $\eta^{\prime}\pi$, $f_{1}\pi$ and $b_{1}\pi$ decay modes appear to be robust in
the observation of a resonant $\pi_{1}(1600)$, there are concerns about the $3\pi$ final 
states. While we report these in the table, the results should be taken with some caution.

\begin{table}[h!]\centering
\begin{tabular}{cllcc} \hline\hline
Mode & Mass (GeV) & Width (GeV) & Experiment & Reference \\ 
\hline
$\rho\pi$            & $1.593\pm 0.08$& $0.168\pm 0.020$ & E852 & \cite{Adams:1998ff} \\
$\eta^{\prime}\pi$ & $1.597\pm 0.010$ & $0.340\pm 0.040$ & E852 & \cite{Ivanov:2001rv} \\
$f_{1}\pi$              & $1.709\pm 0.024$ & $0.403\pm 0.080$ & E852 & \cite{Kuhn:2004en} \\
$b_{1}\pi$             & $1.664\pm 0.008$ & $0.185\pm 0.025$ & E852 & \cite{Lu:2004yn} \\
$b_{1}\pi$             & $1.58\pm 0.03$ & $0.30\pm 0.03$ & VES & \cite{Dorofeev:99} \\
$b_{1}\pi$             & $1.61\pm 0.02$ & $0.290\pm 0.03$ & VES & \cite{Khokholov:00} \\
$b_{1}\pi$             & $\sim 1.6$ & $\sim 0.33$ & VES & \cite{Dorofeev:02} \\
$b_{1}\pi$             & $1.56\pm 0.06$ & $0.34\pm 0.06$ & VES & \cite{Amelin:2005ry} \\
$f_{1}\pi$             & $1.64\pm 0.03$ & $0.24\pm 0.06$ & VES & \cite{Amelin:2005ry} \\
$\eta^{\prime}\pi$  & $1.58\pm 0.03$ & $0.30\pm 0.03$ & VES & \cite{Dorofeev:99} \\
$\eta^{\prime}\pi$  & $1.61\pm 0.02$ & $0.290\pm 0.03$ & VES & \cite{Khokholov:00} \\
$\eta^{\prime}\pi$    & $1.56\pm 0.06$ & $0.34\pm 0.06$ & VES & \cite{Amelin:2005ry} \\
$b_{1}\pi$  & $\sim 1.6$ & $\sim 0.23$ & CBAR & \cite{Baker:2003jh} \\
$\rho\pi$ & $1.660\pm 0.010$ & $0.269\pm 0.021$ & COMPASS & \cite{Alekseev:2009xt} \\
all              & $1.662^{+0.015}_{-0.011}$ & $0.234\pm 0.050$ & PDG & \cite{amsler08} \\
\hline\hline
\end{tabular}
\caption[]{\label{tab:p1_1600}Reported masses and widths of the $\pi_{1}(1600)$ from the 
E852 experiment, the VES experiment and the COMPASS experiment. The PDG average from
2008 is also reported.}
\end{table}

Models for hybrid decays predict rates for the decay of the $\pi_{1}$.
Equation~\ref{eq:decay_rates} gives the predictions from reference~\cite{Close:1995}.
A second model from reference~\cite{page-99} predicted the following rates for a
$\pi_{1}(1600)$.
\begin{center}
\begin{tabular}{lccccc}
      & $\pi b_{1}$ &  $\rho\pi$ & $\pi f_{1}$ &  $\eta(1295)\pi$ & $K^{*}K$  \\ 
PSS & $24$ & $9$ & $5$ & $2$ & $0.8$ \\ 
IKP  & $59$ & $8$ & $14$ & $1$ & $0.4$ 
\end{tabular}
\end{center}
These can be compared to the results from VES in equation~\ref{eq:pi11600_rates3},
which are in moderate agreement. The real identification of the $\pi_{1}(1600)$ as a
hybrid will almost certainly involve the identification of other members of the nonet:
the $\eta_{1}$ and/or the $\eta^{\prime}_{1}$, both of which are expected to have widths
that are similar to the $\pi_{1}$. For the case of the $\eta_{1}$, the most promising 
decay mode may be the $f_{1}\eta$ as it involves reasonably narrow daughters.

We believe that the current data support the existence of a resonant $\pi_{1}(1600)$ which
decays into $b_{1}\pi$, $f_{1}\pi$ and $\eta^{\prime}\pi$, however, near-term confirmation of
these results by COMPASS would be useful. For the $\rho\pi$ decay, we are uncertain. As noted
earlier, the phase motion results observed by both E852 and E852-IU are can be interpreted 
as either the $\pi_{2}(1670)$ absorbing the $\pi_{1}(1600)$, or leakage from the $\pi_{2}(1670)$
generating a spurious signal in the $1^{-+}$ channel. While the new COMPASS result are indeed
interesting, we are concerned about their findings of exactly the same mass and width for 
the $\pi_{2}(1670)$ and the $\pi_{1}(1600)$. We are also concerned that their initial analyses may
be over simplified, particularly in their bias towards an all-resonant description of their data. 
We hope that follow-on results from COMPASS will more broadly explore the model space
imposed by their analyses. We would also like to see results on other final states coupled
to those on three pions. 
\subsection{\label{sec:pi1_2015}The $\pi_{1}(2015)$}
\begin{figure}[h!]\centering
\includegraphics[width=0.35\textwidth,angle=270]{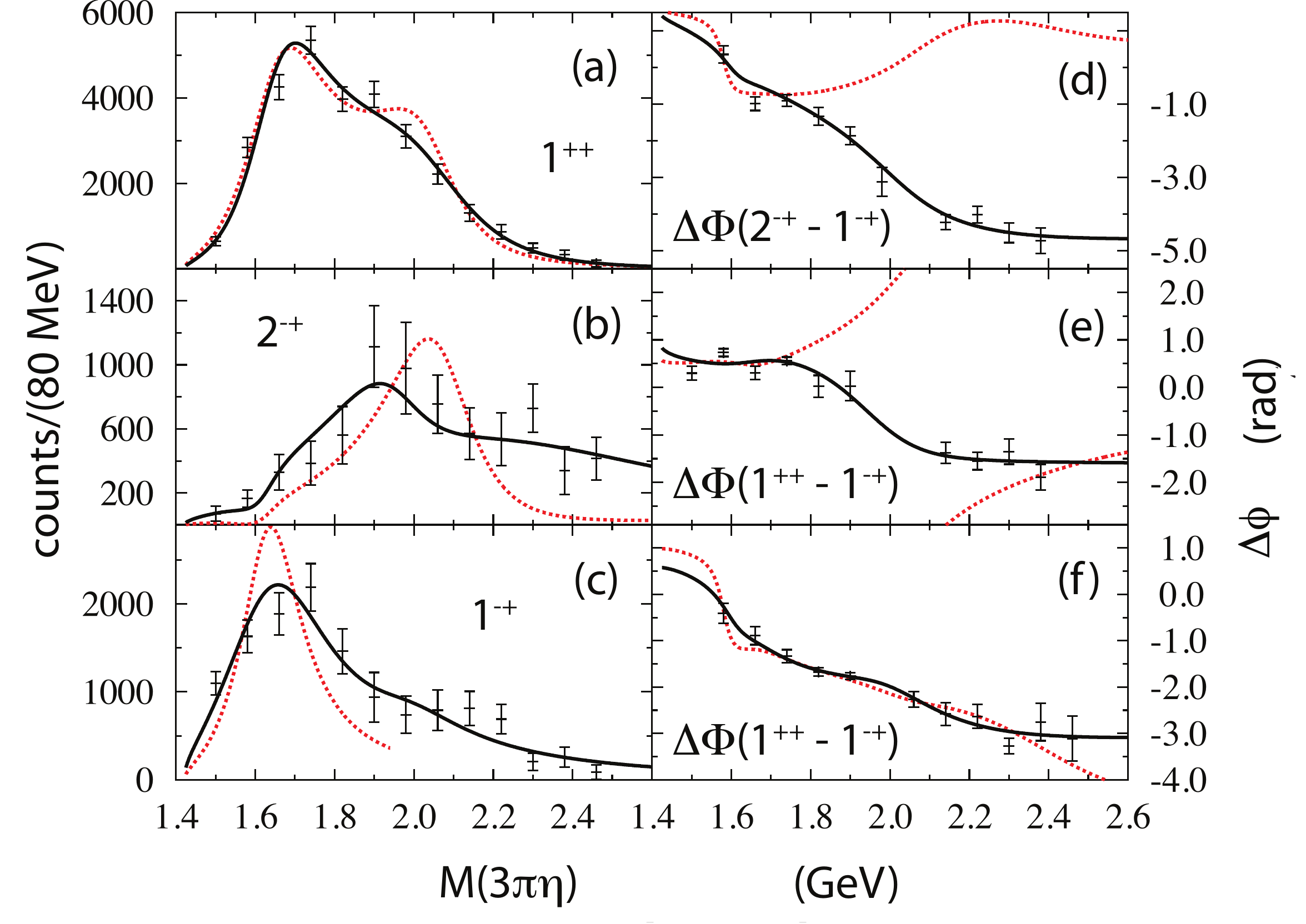}
\caption[]{\label{fig:e852f1pi} (Color on line.) The $f_{1}\pi$ invariant mass from E852~\cite{Kuhn:2004en}.
(a) The $1^{++}$ partial wave ($a_{1}(1270)$), (b) the $2^{-+}$ partial wave ($\pi_{2}(1670)$) and
(c) the exotic $1^{-+}$ partial wave. The dotted (red) curves show the fits of Breit-Wigner
distributions to the partial waves. (d) shows the phase difference between the $2^{-+}$ and 
$1^{-+}$ partial waves, while (e) shows the difference between the $1^{++}$ and $1^{-+}$
partial waves. The dotted (red) curves show the results for a single $\pi_{1}$ state, the $\pi_{1}(1600)$.
(f) shows the same phase difference as in (d), but the dotted (red) curve shows a fit with two
poles in the $1^{-+}$ partial wave, the $\pi_{1}(1600)$ and the $\pi_{1}(2015)$.
(Figure reproduced from reference~\cite{Kuhn:2004en}.)}
\end{figure}

The E852 experiment has also reported a third $\pi_{1}$ state seen decaying to both $f_{1}\pi$
~\cite{Kuhn:2004en} and to $b_{1}\pi$~\cite{Lu:2004yn}. In the $f_{1}\pi$ final state, the
$\pi_{1}(2015)$ is produced with $M^{\epsilon}=1^{+}$ in conjunction with the $\pi_{1}(1600)$.
The description of the $1^{-+}$ partial wave requires two poles. They report a  mass of
$2.001\pm 0.030\pm 0.092$~GeV and a width of $0.333\pm 0.052\pm 0.049$~GeV.
Figure~\ref{fig:e852f1pi} shows the E852 data from this final state. Parts $e$ and $f$ of
this show the need for the two-pole solution.
VES also examined the $f_{1}\pi$ final state, and their intensity of the $1^{-+}$ partial
wave above $1.9$~GeV (see Figure~\ref{fig:ves_f1pi}) is not inconsistent with that of 
E852~\cite{Amelin:2005ry}. However, VES made no comment on this, nor have they 
claimed the existence of the $\pi_{1}(2015)$.

\begin{figure}[h!]\centering
\includegraphics[width=0.5\textwidth]{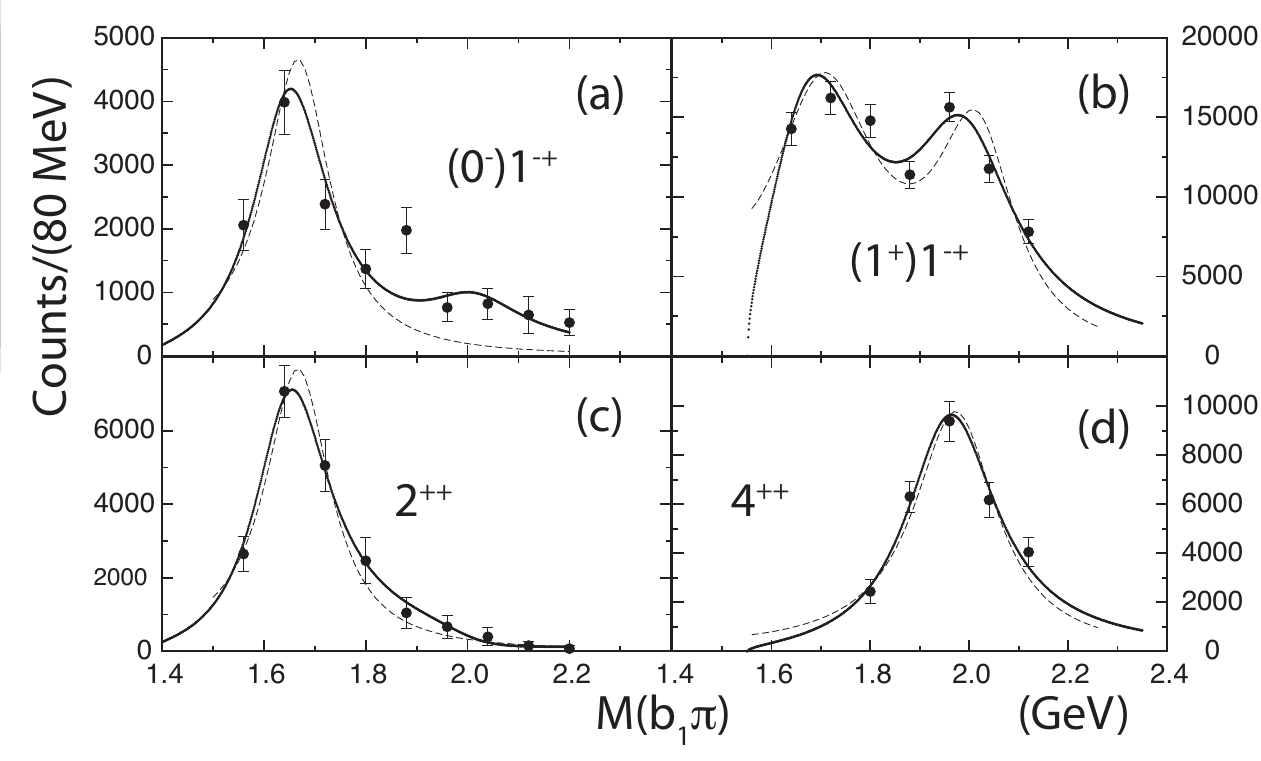}
\caption[]{\label{fig:e852b1pi}
The $b_{1}\pi$ invariant mass from the E852 experiment. (a) shows the $1^{-+}$ $b_{1}\pi$ partial
 wave produced in natural parity exchange ($M^{\epsilon}=1^{+}$) while (b) shows the $1^{-+}$ 
$b_{1}\pi$ partial wave produced in unnatural parity exchange  ($M^{\epsilon}=0^{-}$). In (c) 
is shown the $2^{++}$ $\omega\rho$ partial wave, while (d) shows the $4^{++}$ $\omega\rho$
partial wave. The curves are fits to the $\pi_{1}(1600)$ and $\pi_{1}(2015)$ (a and b), the $a_{2}(1700)$
in (c) and the $a_{4}(2040)$ in (d). (Figure reproduced from reference~\cite{Lu:2004yn}.)}
\end{figure}
In the $b_{1}\pi$ final state, the $\pi_{1}(2015)$ is produced dominantly through natural parity
exchange ($M^{\epsilon}=1^{+}$) while the $\pi_{1}(1600)$ was reported in both natural and 
unnatural parity exchange, where the unnatural exchange dominated. They observe a mass 
of $2.014\pm 0.020\pm 0.016$~GeV  and a width of $0.230\pm 0.032\pm 0.073$~GeV
which are consistent with that observed in the $f_{1}\pi$ final state. Figure~\ref{fig:e852b1pi}
shows the intensity distributions for several partial waves in this final states. The need for
two states is most clearly seen in $b$. VES also looked at the $b_{1}\pi$ final state, but did 
not observe $1^{-+}$ intensity above $1.9$~GeV~\cite{Amelin:2005ry}. However, the intensity
shown in Figure~\ref{fig:ves_b1pi} may be consistent with that observed by E852.
The reported masses and widths are summarized in Table~\ref{tab:p1_2000}. We note that this 
state does not appear in the summary tables of the PDG~\cite{amsler08}. 
\begin{table}[h!]\centering
\begin{tabular}{crrcc} \hline\hline
Mode & Mass (GeV) & Width (GeV) & Experiment & Reference \\ 
\hline
$f_{1}\pi$ & $2.001\pm 0.030$ & $0.333\pm 0.052$ & E852 & \cite{Kuhn:2004en} \\
$b_{1}\pi$ & $2.014\pm 0.020$ & $0.230\pm 0.032$ & E852 & \cite{Lu:2004yn} \\
\hline\hline
\end{tabular}
\caption[]{\label{tab:p1_2000} Reported masses and widths of the $\pi_{1}(2015)$ as observed
in the E852 experiment. The PDG does not report an average for this state.}
\end{table}

With so little experimental evidence for this high-mass state, it is difficult to say much. We note
that the observed decays, $f_{1}\pi$ and $b_{1}\pi$ are those expected for a hybrid meson.
We also note that the production of this state is consistent (natural parity exchange) for both
of the observed final states. In the case that the $\pi_{1}(1600)$ is associated with the lowest-mass
hybrid state, one possible interpretation of the $\pi_{1}(2015)$ would be a excited state (as suggested
by recent LQCD calculations~\cite{Dudek:2010wm}). The mass
splitting is typical of radial excitations observed in the normal mesons. In the case of the 
$\pi_{1}(1600)$ identified as something else, the $\pi_{1}(2015)$ would be a prime candidate 
for the lightest mass hybrid.

\subsection{Other Exotic-quantum Number States}
While no result has been published, the E852 collaboration has presented evidence at conferences
for an isoscalar $2^{+-}$ state~\cite{Adams:2005tx}. The signal is observed with a mass near 
$1.9$~GeV in the $\omega\pi^{-}\pi^{+}$ final state. It decays through $b_{1}\pi$ and is produced 
in both natural and unnatural parity exchange. This conference report was not followed up by a 
publication, so the signal should be viewed with caution. However, if confirmed, this state roughly
lines up in mass with the $\pi_{1}(2015)$ and would be consistent with the lattice picture in
which the $\pi_{1}(1600)$ is the lowest-mass hybrid and the $\pi_{1}(2015)$ is the first 
excitation~\cite{Dudek:2010wm}.

\section{The Future}
The COMPASS experiment has recently started looking at pion peripheral production similar to 
work carried out by both VES and E852. Two new facilities are also expected in the not-too-distant
future, PANDA at GSI and GlueX at Jefferson Lab. The former will study $\bar{p}p$ annihilation 
in the charmonium region, but it will also be possible to search for production of light-quark
hybrids. GlueX will use a $9$~GeV beam of linearly polarized photons to produce hybrids. 

Photoproduction of hybrids is interesting for several reasons. Simple arguments based on vector 
meson dominance suggest that the photon may behave like an $S=1$ $\bar{q}q$ system. 
In several models, such a system is more likely to couple to exotic quantum-number hybrids.
Early calculations of hybrids used the apparent large $\rho\pi$ coupling of the $\pi_{1}(1600)$
to suggest that this state should be produced at least as strongly as normal mesons in
photoproduction~\cite{Afansev-00,Szczepaniak-01,Close-03}. Unfortunately, the current 
controversy on the $\rho\pi$ decay of the $\pi_{1}(1600)$ makes the underlying assumption 
questionable, which may be confirmed by the non observation of the $\pi_{1}(1600)$ by 
CLAS~\cite{Nozar:09}.

Recently, lattice calculations have been performed to compute the radiative decay of 
charmonium $c\bar{c}$ and hybrid states~\cite{Dudek:2009kk}. In the charmonium system,
they find that there is a large radiative decay for an exotic quantum number hybrid. These
studies are currently being extended to the light-quark hybrids with the goal of providing 
estimates of the photoproduction cross sections of these states. However, based on the results
in the charmonium sector, photoproduction appears to be a good place to look for hybrid mesons.

\section{Conclusions}
Over the last two decades, substantial data has been collected looking for 
exotic-quantum-number mesons. In particular, searches have focused on hybrid mesons,
which arise due to excitations of the gluonic fields which confine quarks inside mesons. 
Models and LQCD predictions suggest that three nonets of exotic-quantum-number states
should exist, with $J^{PC}=0^{+-}$,$1^{-+}$ and $2^{+-}$,  where the $1^{-+}$ is expected to be
the lightest. The most recent dynamical calculations of the isovector sector suggest a pair of $1^{-+}$ 
states, with the $0^{+-}$ and $2^{+-}$ states similar in mass to the heavier spin-one state. 
Calculations for the isoscalar states are currently underway, and preliminary results tend to
agree with the isovector spectrum.  Work is also underway to use lighter quark masses. These
masses are measured by quoting the pion mass. Current work has pushed this to $390$~MeV,
and $260$~MeV is in progress. Calculations at the physical pion mass may be within reach.

While not supported by LQCD calculations, other models suggest that exotic-quantum-number 
multiquark states could exist as members of an  $18\oplus\overline{18}$ of SU(3). 
Expected $J^{PC}$ are $1^{-+}$ 
and $0^{--}$, where the spin-one states are expected to be the lightest. The  spin-zero states 
may be similar in mass. However, in order for these multiquark states to have finite widths, some
additional binding mechanism needs to be present to prevent them from simply falling apart
into pairs of mesons. 

Measurements of the $J^{PC}$s, multiplet structure, and decays can be used to distinguish
between these hybrid and multiquark states. However, to do this requires the observation 
of multiple members of a given multiplet as well as observation of states of different $J^{PC}$. 
Experimental results have provided evidence for three $J^{PC}=1^{-+}$ isoscalar states, the 
$\pi_{1}(1400)$, the $\pi_{1}(1600)$ and the $\pi_{1}(2015)$. 

The $\pi_{1}(1400)$ has been observed in both peripheral pion production and $\bar{p}n$
annihilation at rest. It has been seen decaying into $\eta\pi$ (in a p-wave), and even 
though other decay modes have been looked for (such has $\eta^{\prime}\pi$ and $\rho\pi$),
no conclusive evidence for these has been found. While all experiments that have looked at
the $\eta\pi$ final state agree that there is signal strength in the $1^{-+}$ exotic wave, 
the interpretation of this signal is controversial. Explanations exist for the pion production
data that describe the exotic wave as a non-resonant background phase, or produced by
interference with non-resonant processes. Unfortunately, these explanations have not been
tested against the $\bar{p}n$ data.

 If the $\pi_{1}(1400)$ is resonant, it is difficult to 
explain it as a hybrid meson. It mass is too low, and its single decay appears inconsistent
with state being part of an SU(3) nonet. Describing the $\pi_{1}(1400)$ as a multiquark state
is a more natural explanation. However, in reviewing all the experimental evidence, as well 
as the non-resonant descriptions of the $1^{-+}$ signal, we feel that the $\pi_{1}(1400)$
is not resonant.

The most extensive experimental evidence is for the $\pi_{1}(1600)$. It has been observed in
four different decay modes, $\eta^{\prime}\pi$, $b_{1}\pi$, $f_{1}\pi$ and $\rho\pi$, by several 
experiments. Consistent results between E852 and VES are found for the first three decay
modes, and from  $\bar{p}p$ annihilation in flight for the $b_{1}\pi$ mode. However,  the 
$\rho\pi$ decay is controversial. This mode has been observed by two groups, but not 
by two others. In one (VES), the strength is reported in the exotic wave, but they are 
unable to confirm that it is resonant. However, because not all physical constraints 
were used, their conclusions may be weaker than their data would suggest. A second
group, E852-IU, explains the $\pi_{1}(1600)$ as feed through from the stronger 
$\pi_{2}(1670)$ state. However, while the intensity of the $\pi_{1}(1600)$ does depend
on the decays of the $\pi_{2}(1670)$, the phase difference between the two states does not.
This can be interpreted as either feed through from the $\pi_{2}(1670)$, or a resonant
$\pi_{1}(1600)$ being absorbed by the $\pi_{2}(1670)$. To resolve this controversy will likely
require a multi-channel analysis in which physics beyond a simple isobar picture is included.
Even with this controversy about the $\rho\pi$ decay mode, we feel that the experimental 
evidence does support a resonant $\pi_{1}(1600)$. However, confirmation with higher statistics
would be helpful. 

Identification of the $\pi_{1}(1600)$ as the lightest hybrid is not inconsistent with both model 
predictions and LQCD calculations, although some might argue that its mass is somewhat low. 
The current observations and measurements are also consistent with a multiquark interpretation,
although our feeling is that this is less likely. Observation of the isoscalar partners of this
state would help to confirm its hybrid nature. Unfortunately, models predict their 
decays into channels that are experimentally difficult to analyze.

The evidence for the $\pi_{1}(2015)$ is much more limited. It has been seen by one experiment
in two decay modes with very limited statistics while a second experiment (VES) does not see 
evidence for this state. What little is known about this state makes it a good candidate for a hybrid
meson, but confirmation is clearly needed. If both the $\pi_{1}(1600)$ and the $\pi_{1}(2015)$ 
do exist, then the $\pi_{1}(2015)$ may be a radial excitation of the $\pi_{1}(1600)$. A result
which is consistent with the most recent lattice calculations. As with the $\pi_{1}(1600)$, 
observation of the isoscalar partners to this state are important.   

Beyond the  $\eta_{1}$ and $\eta^{\prime}_{1}$ partners of the $\pi_{1}$ states, the crucial missing 
pieces of the hybrid puzzle are the other $J^{PC}$-exotic nonets, $0^{+-}$ and $2^{+-}$. Here,
there is a single hint of an $h_{2}$ state near $1.9$~GeV, but no published results to this effect.
As with the $\eta_{1}$ decays, those of these other nonets are also challenging, and to date, 
all the data that could be used in these searches has come from pion peripheral production.
Definitive observation of these other nonets would provide the missing information to confirm 
the gluonic excitations of QCD. Fortunately, there will soon be four experimental programs 
running (COMPASS at CERN, BES III in Beijing, PANDA at GSI and GlueX at Jefferson Lab) that 
can provide new information on these issues.

\begin{acknowledgments}
The authors would like to thank Jozef Dudek for useful discussions and comments. 
 This work was supported in part by the U.S. Department of Energy under grant No. 
DE-FG02-87ER40315.
\end{acknowledgments}


\end{document}